\documentclass[epsfig,12pt]{article}
\def\beqn{\begin{eqnarray}}
\def\eeqn{\end{eqnarray}}

\def\beq{\begin{equation}}
\def\eeq{\end{equation}}
\def\ba{\beq\new\begin{array}{c}}
\def\ea{\end{array}\eeq}

\newcommand{\gsim}{\lower.7ex\hbox{$
\;\stackrel{\textstyle>}{\sim}\;$}}
\newcommand{\lsim}{\lower.7ex\hbox{$
\;\stackrel{\textstyle<}{\sim}\;$}}

\newcommand{\ntwo}{${\mathcal N}=2\;$}
\newcommand{\ntwot}{${\mathcal N}= \left(2,2\right)\; $}
\newcommand{\ntwoo}{${\mathcal N}= \left(0,2\right)\; $}
\newcommand{\none}{${\mathcal N}=1\;$}

\newcommand{\pt}{\partial}

\renewcommand{\theequation}{\thesection.\arabic{equation}}

\begin{document}


\begin{titlepage}

\begin{flushright}
FTPI-MINN-08/04, UMN-TH-2534/08\\
ITEP-TH-09/08\\
March 2, 2008
\end{flushright}

\begin{center}

{\Large \bf   Heterotic Flux Tubes in \boldmath\ntwo SQCD\\ [1mm]
with \boldmath\none Preserving Deformations }
\end{center}

\begin{center}
{\bf M.~Shifman$^{a}$ and \bf A.~Yung$^{a,b,c}$}
\end {center}
\vspace{0.3cm}
\begin{center}

$^a${\it  William I. Fine Theoretical Physics Institute,
University of Minnesota,
Minneapolis, MN 55455, USA}\\
$^{b}${\it Petersburg Nuclear Physics Institute, Gatchina, St. Petersburg
188300, Russia}\\
$^c${\it Institute of Theoretical and Experimental Physics, Moscow
117259, Russia}
\end{center}

\begin{abstract}

We consider non-Abelian BPS-saturated flux tubes (strings)
in \ntwo super\-symmetric QCD deformed by
superpotential terms of a special type breaking
\ntwo supersymmetry down to \none$\!\!$.
Previously it was believed that
worldsheet supersymmetry is ``accidentally" enhanced
due to the facts that ${\mathcal N} =(1,1)$ SUSY
is automatically elevated up to \ntwot on $CP(N-1)$ and, at the same time,
there are no \ntwoo generalizations of the bosonic $CP(N-1)$ model. 
Edalati and Tong noted that the target space is in fact
$CP(N-1)\times C$ rather than $CP(N-1)$.
This allowed them to suggest a ``heterotic" \ntwoo 
sigma model, with the $CP(N-1)$ target space for bosonic fields and an extra 
right-handed fermion which couples to the fermion fields of the
\ntwot $CP(N-1)$ model. We derive the heterotic  \ntwoo worldsheet
model directly from the bulk theory. The relation between the bulk and worldsheet deformation parameters we obtain does not coincide with that suggested by Edalati and Tong at large values of the deformation parameter. For 
polynomial deformation superpotentials in the bulk we find nonpolynomial
response in the worldsheet model. We find a geometric representation for the
heterotic model. Supersymmetry is proven to be spontaneously broken for small
deformations (at the quantum level). This confirms Tong's conjecture. A proof valid for large deformations will be presented in the subsequent publication.

\end{abstract}

\end{titlepage}

\newpage
{\small \tableofcontents}
\newpage

\section{Introduction}
\label{intro}
\setcounter{equation}{0}

Non-Abelian BPS-saturated flux tubes were discovered and studied in \ntwo
super\-symmetric QCD \cite{HT1,ABEKY,SYmon,Tong,HT2}.
The simplest model supporting such flux tubes,
to be referred to as the basic model, has the gauge group
U$(N)$, with the U(1) Fayet--Iliopoulos (FI) term,
and $N$ flavors ($N$ hypermultiplets in the fundamental representation).
Multiple developments in supersymmetric solitons and ideas about confinement ensued
(for reviews see \cite{Trev,Etor,SYrev}).
A crucial feature of  non-Abelian strings
is the presence of orientational (and superorientational) moduli associated with
rotations of their color fluxes inside a non-Abelian group, in addition to
``standard" translational and supertranslational moduli. The low-energy theory on the
string worldsheet is split into two disconnected parts:
a free theory for (super)translational moduli and a nontrivial part, a theory of
interacting (super)orientational moduli, $CP(N-1)$ model.
 The latter is completely fixed by the fact
that the basic bulk theory has eight supercharges, and the string under consideration
is 1/2 BPS. As well-known (e.g. \cite{BW,NSVZsigma}), the only supergeneralization of the bosonic $CP(N-1)$ is the \ntwot  supersymmetric $CP(N-1)$ model with four supercharges.

In a bid to decrease the level of supersymmetry (SUSY) in the bulk theory
an \ntwo breaking deformation of the type 
\beq
{\mathcal W}_{\rm deform} = \mu {\mathcal A}^2
\label{defo}
\eeq
was introduced \cite{SYnone} where ${\mathcal A}$ is the adjoint chiral superfield.
The above deformation preserves \none in the bulk. As $\mu$ increases, the adjoint fields become heavier and eventually decouple from the spectrum at $\mu\to \infty$.

With \none preserving deformation of the basic model, there are 
four conserved supercharges in the bulk rather than eight. At the same time, 
the description of the orientational moduli is the same as in the \ntwo
basic model; the bosonic part of the worldsheet theory is $CP(N-1)$.
Since the string solution
remains 1/2 BPS, the worldsheet theory must have two conserved supercharges.
Endowing the bosonic $CP(N-1)$ model with two supercharges
automatically endows it with four supercharges \cite{BW,NSVZsigma}. A conclusion was made \cite{SYnone} that in the problem at hand,
unexpectedly, the worldsheet supersymmetry enhances up to \ntwot$\!.$
If it were the case, the situation would be similar to
supersymmetry enhancement on domain walls \cite{Ritz}.

Recently Edalati and Tong noted \cite{Edalati} that
the bosonic part of the worldsheet sigma model on the string is, in fact,  
$CP(N-1)\times C$ rather than $CP(N-1)$, and endowing $CP(N-1)\times C$ with two supercharges need not necessarily lead to \ntwot supersymmetry on the worldsheet. They built an \ntwoo heterotic model which supergeneralizes the bosonic model with the above target space. Moreover, basing on a number of indirect checks
they concluded that the Edalati--Tong heterotic model emerges on the
string worldsheet in the \none bulk theory and suggested a rule of converting
the bulk \ntwo breaking superpotential into an \ntwot
breaking superpotential on the string worldsheet.

To be more exact, the Edalati--Tong model is designed as follows.
Consider for example the U(2) model in the bulk with
$CP(1)$ on the worldsheet. If \ntwo in the bulk is unbroken,
the 1/2 BPS flux tube has two translational moduli associated with its center $x_0$,
and four supertranslational moduli. The above set is totally decoupled from
two orientational moduli parameterizing the coset SU(2)/U(1) accompanied
by four superorientational moduli. 

When \ntwo is broken by an \none$\!$-preserving deformation,
the number of the moduli fields remains intact, but their grouping
changes. The four supertranslational moduli split into two plus two.
Two left-handed fermion fields combine with $x_0$
to form an \ntwoo supermultiplet. These fields are described by a free theory and are decoupled from the rest of the worldsheet theory. (General aspects of two-dimensional
\ntwoo sigma models were discussed in \cite{EWI}.)

The right-handed fermion fields $\zeta_R$ and $\bar\zeta_R$, which used to be
``two other" supertranslational moduli, ``mix" with two
right-handed superorientational moduli
tangential to the  coset SU(2)/U(1). Together with two orientational 
moduli of $CP(1)$ and four superorientational moduli
they form the \ntwoo extension of the $CP(1)$ model. 
For brevity sometimes we will refer to it as the heterotic $CP(1)$ (or heterotic 
$CP(N-1)$ for generic $N$). The 
fermion fields $\zeta_R$ and $\bar\zeta_R$ lie outside
the target space SU(2)/U(1). They are remnants of $C$. With respect to
\ntwoo supersym\-metry they transform through $F_\zeta$ terms which
are expressible, via equations of motion, 
in terms of the fermion fields of the conventional $CP(1)$ model. 

In this paper we present a {\em direct derivation} of the string worldsheet theory
for a generic superpotential in the bulk theory breaking \ntwo while preserving
\none {\em and} the 1/2-BPS nature of the flux tube solution at the classical level. 
The $\mu {\mathcal A}^2$ superpotential mentioned above is a particular case.
Generally speaking, the minimal choice one can consider is a cubic in ${\mathcal A}$
superpotential (in  the U(2) bulk theory) with coefficients rigidly fixed
by the quark mass terms. In the U$(N)$ bulk theory with $N_f=N$ flavors
the minimal
admissible \none$\!$-preserving deformation is a polynomial of the $(N+1)$-th order
whose coefficients are unambiguously fixed. These more general superpotentials will be considered as well.

Focusing on the simplest example of U(2) in the bulk 
we prove that an \ntwoo extension of the
$CP(1)$ model \`a la Edalati--Tong does indeed emerge on the string worldsheet in the low-energy limit. While gross features of the emergent heterotic worldsheet
theory are those predicted by Edalati and Tong,
details do not quite coincide. In particular, for polynomial deformations 
in the bulk we find, generally speaking, a non-polynomial
response in the worldsheet theory. Our direct derivation of the heterotic string model relies,
in addition to already known results, on explicit form of the fermion zero 
modes on the BPS flux tubes in \none bulk theories. To obtain the fermion 
zero modes we had
to extend previous analyses \cite{SYnone,VY}. Thus, a large part of this paper bears a technical nature. It is based, however, on an observation 
of conceptual nature (Sect. \ref{02})
which is responsible for the very possibility of direct derivation
of the heterotic $CP(1)$ model on the string worldsheet. Indeed, in the 
Edalati--Tong formulation the difference between the
\ntwot and heterotic models reveals itself in four-fermion terms. It is very hard,
if possible at all, to derive these terms starting directly from the bulk theory.
In our formulation the most straightforward distinction between two models
occurs in the kinetic part of the Lagrangian, in the term {\em bilinear} in the fermion fields,
of the type
\beq
\left(\zeta_R^\dagger \chi^a_R\right) \partial_L \, S^a\,,
\label{bilin}
\eeq
where $S^a$ is the bosonic field of the $O(3)$ model subject to the constraint
$\vec S^{\,2}=1$, while $\chi^a_R$ is its fermionic superpartner,
$\vec S \vec\chi =0$. Since the term in (\ref{bilin})
is bilinear in the fermion fields, the knowledge of the fermion zero modes
allows one to get this term from the bulk Lagrangian in a very explicit and direct way.
Other additional terms transforming \ntwot model into \ntwoo
unambiguously follow from (\ref{bilin}) by virtue of \ntwoo supersymmetry.

The basic features of the heterotic $CP(1)$ model we obtain are as follows.
The term (\ref{bilin}) entails  the
occurrence of the four-fermion interaction of the type
\beq
\left(\zeta_R^\dagger \zeta_R\right)\left(\chi_L^{a}\chi_L^b\right)
S^c\,\varepsilon_{abc}\,,
\eeq
and a suppression of the coefficient in front
of the conventional four-fermion term
\beq
\frac{1}{2}\, \left(\chi_L^{a}\chi_R^a\right)^2\,.
\label{cff}
\eeq

\vspace{3mm}

The addition of seemingly rather insignificant
$\zeta_R\,, \bar\zeta_R$ terms to the \ntwot \- $CP(N-1)$ model
drastically changes its dynamical behavior. In particular, Witten's index
$I=N$ for $CP(N-1)$ \cite{WI} changes and becomes zero. 
Supersymmetry on the worldsheet is no longer protected 
by Witten's index. In fact, we will prove, at small $\mu$, that spontaneous SUSY breaking does take place. The fields $\zeta_R,\,\zeta^\dagger_R$
play the role of Goldstinos.
In the accompanying paper \cite{ACC}
we will solve the heterotic $CP(N-1)$ model at large $N$
and prove that supersymmetry is spontaneously broken at the quantum level
for any value of the deformation parameter,
as was anticipated by Tong \cite{Tongd}.  This result seems to be intuitively clear given
that small variations of the deformation superpotential
ruin the BPS nature of the flux-tube solutions already at the classical level.

We will derive a long-sought geometric representation of the
heterotic \ntwoo model, in terms of the metric and curvature tensor
of the $CP(N-1)$ space. 

\vspace{3mm}

Organization of the paper is as follows. In Sect.~\ref{bulk}
we review our basic bulk theory with eight supercharges
and discuss possible deformations of this bulk theory breaking \ntwo down to \none
without destroying the BPS nature of the flux-tube solution. In Sect.~\ref{strings}
we review construction of non-Abelian strings in the \ntwot limit.
Moreover, we perform derivation of those fermion zero modes
which had not been explicitly derived in the literature previously.
Section~\ref{0,2} summarizes general aspects of the Edalati--Tong model.
In Sect.~\ref{02} we present our formulation of the heterotic $CP(1)$
model. Section~\ref{geom} is devoted to yet another, geometric, formulation
of  the heterotic $CP(1)$ model. 
Here we also show that at small $\mu$ the vacuum energy density of the
heterotic model is proportional to the square of the chiral condensate.
In Sect.~\ref{N1zeromodes}
we begin our direct derivation of the worldsheet model
from the bulk theory deformed by the superpotential (\ref{defo}). 
Section~\ref{N1zeromodes} is devoted 
to the fermion zero modes. Section~\ref{crossterm}
establishes the relation between the parameters of the worldsheet model
and those of the bulk theory. In Sects.~\ref{twistedmass} and \ref{deformationsup}
we proceed to a more general case of a polynomial deformation superpotential
replacing the simplest superpotential (\ref{defo}). Here we calculate the
worldsheet superpotential in two limits, $\mu\to 0$ and $\mu\to\infty$.
While the first result agrees with the Edalati--Tong conjecture,
the large-$\mu$ limit defies it. We show that in this case the main effect of \ntwo breaking deformation
in the $\mu\to\infty$ limit is that the potential of the worldsheet theory gets
enhanced. It forces the string orientational vector to point towards the 
north or south poles of 
the sphere $S_2={\rm SU}(2)/{\rm U}(1)$. The string becomes exceedingly more ``Abelian'' as we increase the deformation superpotential in the bulk.
Section~\ref{conclu} summarizes our findings. 

\vspace{1mm}
{\bf Remark:} In Sects.~\ref{bulk}--\ref{02} and \ref{N1zeromodes}--\ref{deformationsup} we use Euclidean notation most suitable for
consideration of static solitons. This is explained in Appendix A.
Section~\ref{geom} which bears a general nature
is presented in Minkowski notation. This is explained in Appendix B.
In Appendix C we briefly discuss the Witten index for the
heterotic \ntwoo $CP(N-1)$ models. In Appendix D we collect for convenience various
definitions of the deformation parameters.

\section{Bulk theory}
\label{bulk}
\setcounter{equation}{0}

The gauge symmetry of the basic bulk model is 
SU$(N)\times$U(1). We will focus on the SU(2)$\times$U(1) case,
which presents the simplest example.
Besides the gauge bosons, gauginos and their  superpartners,
the model has the matter sector consisting of $N_f=N=2$ ``quark" hypermultiplets. 
In addition, we will introduce the  Fayet--Iliopoulos $D$-term for the U(1) gauge field
which triggers the quark condensation.

Let us first discuss the undeformed theory with
 \ntwo$\!.$ The  superpotential has the form
 \beq
{\mathcal W}_{{\mathcal N}=2} =\frac{1}{\sqrt 2} \sum_{A=1}^2
\left( \tilde q_A {\mathcal A}
q^A +  \tilde q_A {\mathcal A}^a\,\tau^a  q^A\right)\,,
\label{superpot}
\eeq
where ${\mathcal A}$ and ${\mathcal A}^a$ are  chiral superfields, the ${\mathcal N}=2$
superpartners of the gauge bosons of  U(1) and SU(2), respectively.
Furthermore, $q_A$ and $\tilde q_A$ ($A=1,2$) represent two
matter (quark) hypermultiplets. The flavor index is denoted by $A$. Thus,
in our model the number of colors coincides with the number of flavors. The $q^A$ mass
terms are denoted by $m_A$.

Next, we add a superpotential which breaks
supersymmetry down to \none$\!.$ In this paper we will consider two types of 
\none preserving deformation superpotentials. The first 
superpotential is the mass term for the adjoint fields,
\beq
{\mathcal W}_{3+1}=\frac{\mu}{2} \left[{\mathcal A}^2
+  ({\mathcal A}^a)^2\right],
\label{msuperpotbr}
\eeq
where $\mu$ is a common mass parameter for the chiral
superfields in \ntwo gauge supermultiplets,
U(1) and SU(2), respectively. The subscript 3+1 tells us that the deformation 
superpotential (\ref{msuperpotbr}) refers to the bulk four-dimensional theory.
Clearly, the mass term (\ref{msuperpotbr}) splits \ntwo supermultiplets, breaking
\ntwo supersymmetry down to \none.

For the deformation (\ref{msuperpotbr}), in order to preserve the BPS nature of the flux-tube solutions, it is necessary to set the quark mass terms at zero,
\beq
\label{zeromass}
m_1=m_2=0\,.
\eeq
As was shown in \cite{SYnone} and \cite{Edalati} (see also the review 
paper \cite{SYrev}), in this case the deformed theory supports 1/2 BPS -saturated
flux-tube solutions at the classical level.

The second (more general) deformation we will consider in this paper is a   polynomial
superpotential of the form 
\beq
\label{brsup}
{\mathcal W}_{3+1}={\rm Tr}\,\sum_{k=1}^{N=2}\frac{c_k}{k+1}\,\hat{{\mathcal A}}^{k+1}\,,
\eeq
where we introduce the adjoint matrix superfield
\beq
\hat{{\mathcal A}}=\frac12 {\mathcal A}+\frac{\tau^a}{2} {\mathcal A}^a\,;
\label{admatrix}
\eeq
$\tau^a$ are the SU(2) Pauli matrices. The hat over ${\mathcal A}$
will remind us that ${\mathcal A}$ is a matrix from U(2) rather than SU(2).
The coefficients $c_k$ are not arbitrary. As explained at the end of this section, they are unambiguously fixed by the bulk theory parameters.
 
The bosonic part of our SU(2)$\times$U(1) theory has the form
\beqn
S&=&\int d^4x \left[\frac1{4g^2_2}
\left(F^{a}_{\mu\nu}\right)^2 +
\frac1{4g^2_1}\left(F_{\mu\nu}\right)^2
+
\frac1{g^2_2}\left|D_{\mu}a^a\right|^2 +\frac1{g^2_1}
\left|\partial_{\mu}a\right|^2 \right.
\nonumber\\[4mm]
&+&\left. \left|\nabla_{\mu}
q^{A}\right|^2 + \left|\nabla_{\mu} \bar{\tilde{q}}^{A}\right|^2
+V(q^A,\tilde{q}_A,a^a,a)\right]\,.
\label{model}
\eeqn
Here $D_{\mu}$ is the covariant derivative in the adjoint representation
of  SU(2),
while
\beq
\nabla_\mu=\partial_\mu -\frac{i}{2}\; A_{\mu}
-i A^{a}_{\mu}\, \frac{\tau^a}{2}.
\label{defnabla}
\eeq
The coupling constants $g_1$ and $g_2$
correspond to the U(1)  and  SU(2)  sectors  respectively.
With our conventions, the U(1) charges of the fundamental matter fields are 
$\pm 1/2$.

\vspace{1mm}

The potential $V(q^A,\tilde{q}_A,a^a,a)$ in the Lagrangian (\ref{model})
is a sum of  various $D$ and  $F$  terms,
\beqn
V(q^A,\tilde{q}_A,a^a,a) &=&
 \frac{g^2_2}{2}
\left( \frac{1}{g^2_2}\,  \varepsilon^{abc} \bar a^b a^c
 +
 \bar{q}_A\,\frac{\tau^a}{2} q^A -
\tilde{q}_A \frac{\tau^a}{2}\,\bar{\tilde{q}}^A\right)^2
\nonumber\\[3mm]
&+& \frac{g^2_1}{8}
\left(\bar{q}_A q^A - \tilde{q}_A \bar{\tilde{q}}^A-2\xi\right)^2
\nonumber\\[3mm]
&+& \frac{g^2_2}{2}\left| \tilde{q}_A\tau^a q^A 
+\sqrt{2}\,\,\frac{\pt{\mathcal W}_{3+1}}{\pt a^a}
\right|^2+
\frac{g^2_1}{2}\left| \tilde{q}_A q^A +
\sqrt{2}\,\,\frac{\pt{\mathcal W}_{3+1}}{\pt a}
 \right|^2
\nonumber\\[3mm]
&+&\frac12\sum_{A=1}^2 \left\{ \left|(a +\tau^a a^a +\sqrt{2}m^A)q^A
\right|^2 
\right.
\nonumber\\[3mm]
&+&
\left.
\left|(a +\tau^a a^a +\sqrt{2}m^A)\bar{\tilde{q}}_A
\right|^2 \right\}\,,
\label{pot}
\eeqn
where the sum over repeated flavor indices $A$ is implied.
The first and second lines here represent   $D$   terms, the third line
the $F_{\mathcal A}$ terms,
while the fourth and the fifth lines represent the squark $F$ terms.
We also introduced the Fayet--Iliopoulos  $D$-term for the U(1) field,
with the FI parameter $\xi$ in (\ref{pot}).
Note that the Fayet--Iliopoulos term does not
break \ntwo supersymmetry \cite{matt,VY}. The parameters which do
break  \ntwo   down to \none are $\mu$ or $c_k$ in (\ref{msuperpotbr}) or
(\ref{brsup}).

The vacuum structure and the mass spectrum of perturbative excitations
in this theory were studied in \cite{SYnone} for the case of mass-type deformation 
(\ref{msuperpotbr}). Here we briefly review relevant results for convenience.

The Fayet--Iliopoulos term triggers the spontaneous breaking
of the gauge symmetry. The vacuum expectation values (VEV's)
of the squark fields can be chosen as
\beqn
\langle q^{kA}\rangle &=&\sqrt{
\xi}\, \left(
\begin{array}{cc}
1 & 0 \\
0 & 1\\
\end{array}
\right),\,\,\,\langle \bar{\tilde{q}}^{kA}\rangle =0,
\nonumber\\[3mm]
k&=&1,2,\qquad A=1,2\,,
\label{qvev}
\eeqn
while the VEV's of the adjoint fields vanish
\beq
\langle a^a\rangle =0,\,\,\,\,\langle a\rangle =0.
\label{avev}
\eeq
Here we write down  $q$ as a $2\times 2$ matrix, the first
superscript ($k=1,2$) refers to SU(2) color, while the second
($A=1,2$) to flavor. We keep the quark masses $m_1=m_2=0$ in conjunction with (\ref{msuperpotbr}).

The color-flavor locked form of the quark VEV's in
Eq.~(\ref{qvev}) and the absence of VEV of the adjoint scalar $a^a$ in
Eq.~(\ref{avev}) results in the fact that, while the theory is fully Higgsed, a diagonal
SU(2)$_{C+F}$ survives as a global symmetry.
The presence of this symmetry leads to the emergence of
orientational zero modes of $Z_2$ strings in the model (\ref{model})
\cite{ABEKY}.

With two matter hypermultiplets, the  SU(2) part of the gauge group
is asymptotically free,  implying generation of a dynamical scale $\Lambda$.
In order to stay at weak coupling   we assume
that $\sqrt{\xi}\gg \Lambda$, so that the SU(2) coupling running is 
frozen by the squark condensation at a small value. 

Since both U(1) and SU(2) gauge groups are broken by the squark condensation,
all gauge bosons become massive. From (\ref{model}) we get for the U(1)
gauge boson
\beq
m_{\gamma}=g_1\sqrt{\xi}\,,
\label{phmass}
\eeq
while   three gauge bosons of the SU(2) group acquire the same mass
\beq
m_{W}=g_2\sqrt{\xi}\,.
\label{wmass}
\eeq

To get the masses of the scalar bosons we expand the potential (\ref{pot})
near the vacuum (\ref{qvev}), (\ref{avev}) and diagonalize the
corresponding mass matrix. The four components of the
eight-component\,\footnote{We mean here eight {\em real} components.}
scalar $q^{kA}$ are eaten by the Higgs mechanism for U(1) and SU(2)
gauge groups. Another four components are split as follows:
one component acquires the mass (\ref{phmass}). It becomes
a scalar component of  a massive \none vector U(1) gauge multiplet.
Other three components acquire masses (\ref{wmass}) and become
scalar superpartners of the SU(2) gauge boson in \none massive gauge
supermultiplet.

Other 16 real scalar components of the fields $\tilde{q}_{Ak}$, $a^a$ and $a$
produce the following states: two states acquire mass
\beq
m_{{\rm U}(1)}^{+}=g_1\sqrt{\xi\lambda_1^{+}}\,,
\label{u1m1}
\eeq
while the mass of other two states is given by
\beq
m_{{\rm U}(1)}^{-}=g_1\sqrt{\xi\lambda_1^{-}}\,,
\label{u1m2}
\eeq
where $\lambda_1^{\pm}$ are two roots of the quadratic equation
\beq
\lambda_i^2-\lambda_i(2+\omega^2_i) +1=0\,,
\label{queq}
\eeq
for $i=1$. Here we introduced two \ntwo supersymmetry breaking
parameters associated with the U(1) and SU(2) gauge groups, respectively,
\beq
\omega_1=\frac{g^2_1\mu}{m_{\gamma}}\, ,\qquad
\omega_2=\frac{g^2_2\mu}{m_W}\,.
\label{omega}
\eeq
Furthermore,
other 2$\times$3=6 states acquire mass
\beq
m_{{\rm SU}(2)}^{+}=g_2\sqrt{\xi\lambda_2^{+}}\,,
\label{su2m1}
\eeq
while the remaining 2$\times$3=6 states also become massive. Their mass is
\beq
m_{{\rm SU}(2)}^{-}=g_2\sqrt{\xi\lambda_2^{-}}\,.
\label{su2m2}
\eeq
Here $\lambda_2^{\pm}$ are two roots of the quadratic equation
(\ref{queq}) for $i=2$. Note that all states come either as  singlets
or triplets of unbroken SU(2)$_{C+F}$.

In the large-$\mu$ limit
the larger masses $m_{{\rm U}(1)}^{+}$ and $m_{{\rm SU}(2)}^{+}$ become
\beq
m_{{\rm U}(1)}^{+}= m_{{\rm U}(1)}\omega_1=g_1^2\mu\,,\qquad
m_{{\rm SU}(2)}^{+}= m_{{\rm SU}(2)}\omega_2=g_2^2\mu\, .
\label{amass}
\eeq
Clearly, in the limit $\mu\to \infty$ these are
the masses of the heavy adjoint
scalars $a$ and $a^a$. At $\omega_i\gg 1$ these fields decouple and
can be integrated out.

The low-energy bulk theory in this limit
contains massive gauge  \none multiplets and chiral multiplets with
lower masses $m^{-}_{{\rm U}(1),{\rm SU}(2)}$. Equation (\ref{queq}) gives for these masses
\beq
m_{{\rm U}(1)}^{-}= \frac{m_{{\rm U}(1)}}{\omega_1}
=
\frac{\xi}{\mu}\,,\qquad
m_{{\rm SU}(2)}^{-}= \frac{m_{{\rm SU}(2)}}{\omega_2}
=
\frac{\xi}{\mu}\,.
\label{light}
\eeq
In  the limit of infinite $\mu$ these masses tend to zero.
This fact reflects the emergence of a Higgs branch in \none SQCD,
see, for example, \cite{IS}.

As was explained in \cite{SYnone}, the presence of the Higgs branch in the 
$\mu\to \infty$ limit is quite  an unpleasant feature of the theory (\ref{model}).
The presence of quark massless states in the bulk
associated with this Higgs branch obscure physics of the non-Abelian strings
in this theory. In particular, the strings become infinitely thick.
This means that higher derivative corrections in the effective theory on the string become
important.
In \cite{SYnone} the maximal critical value of the parameter $\mu$ was estimated
beyond which one can no longer trust the effective low-energy theory on 
the string worldsheet,
\beq
g^2_2\mu \ll \frac{m_W^{3}}{\Lambda^2_{{\mathcal N}=1}},
\label{limit}
\eeq
where $\Lambda_{{\mathcal N}=1}$ is the scale of  \none SQCD to which the theory
(\ref{model}) flows in the large-$\mu$ limit,
\beq
\Lambda^4_{{\mathcal N}=1}=g_2^4\mu^2\Lambda^2.
\label{lambdaN=1}
\eeq
We assume  that the condition (\ref{limit}) is met.

We still have a large window for the values of the  $\mu$ parameter,
with $\mu$ staying below the upper bound (\ref{limit}), but, on the other hand, 
large enough to ensure the decoupling of the adjoint fields, 
namely\,\footnote{When we speak 
of sending $\mu$ to $\infty$ we in fact mean that $\mu$ lies near the upper edge of the window (\ref{window}). The dimensionless parameter determining whether
$\mu$ is small or large is $g^2_2\mu/m_W$. When $\mu$ is close to
the upper edge of the window (\ref{window}) for all practical purposes we can put
the above parameter to $\infty$.
}
\beq
m_W\ll g^2_2\mu \ll m_W\, \frac{m_W^{2}}{\Lambda^2_{{\mathcal N}=1}}\,.
\label{window}
\eeq

To conclude this section we briefly discuss a more general deformation of \ntwo
SQCD given by the superpotential (\ref{brsup}). As was shown in \cite{Edalati}, in order to
preserve the BPS nature of the string solutions, one has to consider a deformation 
superpotential (\ref{brsup}) of a special type,  with the critical points coinciding with the quark mass terms. In the U(2) case this boils down to 
\beq
\frac{\pt{\mathcal W}_{3+1}}{\pt\hat{{\mathcal A}}} ={\rm Tr}\,\sum_{k=1}^{N=2}
c_k\,\hat{{\mathcal A}}^{k} =
\frac{\mu}{\Delta m}\, {\rm Tr}\,
\left(\hat{{\mathcal A}}+\frac{m_1}{\sqrt{2}}\right)
\left(\hat{{\mathcal A}}+\frac{m_2}{\sqrt{2}}\right),
\label{defsup}
\eeq
where $\mu$ is the deformation parameter and 
\beq
\Delta m=m_1-m_2\,.
\label{deltam}
\eeq

With this superpotential added, the squark VEV's are given by the same expression
(\ref{qvev}) as for the adjoint mass deformation (\ref{msuperpotbr}), while the adjoint VEV's are now
\beq
\hat{a}=-\frac{1}{\sqrt{2}}\, \left(
\begin{array}{cc}
m_1 & 0 \\
0 & m_2\\
\end{array}
\right)\,.
\label{avevm}
\eeq
It is rather obvious that deviations of the coefficients $c_k$
from (\ref{defsup}) eliminate BPS saturated flux-tube solutions, see Sect.~\ref{strings}.

The deformation (\ref{defsup}) in the large-$\mu$ limit gives large masses,
of the order of $g^2\mu$, to the adjoint fields $a$ and $a^3$ leaving 
the fields $a^{1,2}$ intact (with masses of the order of $g\sqrt{\xi}$,
we assume that $\Delta m\ll g\sqrt{\xi}$). Thus,
in the large-$\mu$ limit
the breaking of the U(2) gauge group by the adjoint VEV's is not washed out.
Instead, it becomes stronger as we increase $\mu$. The 
theory with the deformation (\ref{defsup}) does not flow to \none SQCD
at $\mu\to\infty$ because $m(a^{1,2})$ stays frozen at $g\sqrt \xi$.
In this sense, the mass-type deformation (\ref{msuperpotbr})
is more efficient.

Below in Sects.~\ref{strings}--\ref{crossterm} we discuss the deformation
of \ntwo SQCD with the monomial superpotential (\ref{msuperpotbr}), and then in Sects.~\ref{twistedmass} and \ref{deformationsup} consider the polynomial deformation superpotential (\ref{defsup}).

\section{Non-Abelian strings}
\label{strings}
\setcounter{equation}{0}

Non-Abelian strings were shown to emerge at weak coupling in
\ntwo  supersymmetric gauge theories with the U$(N)$ gauge group \cite{HT1,ABEKY,SYmon,HT2}, see also the review papers \cite{Trev,SYrev}.
The main feature of the non-Abelian strings is the
presence of orientational zero modes associated with rotations of their
color flux in the non-Abelian gauge group, which makes such strings
genuinely non-Abelian. This solution of the \ntwo
theory was generalized to the theory with the mass term deformation 
(\ref{msuperpotbr}) in \cite{SYnone}. 
The reason why the string solution remains BPS-saturated (at the classical level)
after the deformation (\ref{msuperpotbr}) is switched on is as follows:
classically the flux-tube solution is 
constructed from the gauge and $q$ fields which have the same masses, see Eqs.
(\ref{wmass}) and (\ref{phmass}). The fields $\tilde{q}$ and $a$ are given by their
vanishing VEV's, see (\ref{qvev}) and (\ref{avev}). If we considered 
more generic deformations (say, a
deformation of the type (\ref{brsup}) with polynomial superpotentials) 
the fields $\tilde{q}$ and $a$ would be excited in the flux-tube solution implying
the loss of the BPS saturation. The reason is that 
the fields $\tilde{q}$ and $a$ have masses different from those
of the gauge bosons and $q$ fields. 

Below in this and subsequent sections we will consider the mass term deformation 
(\ref{msuperpotbr}) which does not excite the fields $\tilde{q}$, and $a$, and the string
remains classically BPS-saturated.\footnote{The $\tilde{q}$ and $a$ fields are, 
of course, present at the quantum 
level. This raises the issue of a possible breaking of the string ``BPSness" 
at the quantum level (a spontaneous
supersymmetry breaking in the effective theory on the string worldsheet). In \cite{Tongd}
it was argued that \ntwoo supersymmetry in the two-dimensional theory on the string worldsheet is spontaneously broken, indeed. This argument is confirmed
in Sect.~\ref{geom} and by the
exact solution at large $N$, see \cite{ACC}.}

The $Z_2$ string solution
(a progenitor of the non-Abelian string) can be written as
follows \cite{ABEKY}:
\beqn
q(x)
&=&
\left(
\begin{array}{cc}
e^{ i \, \alpha  }\phi_1(r) & 0  \\
0 &  \phi_2(r) \\
\end{array}\right),
\nonumber\\[4mm]
A^3_{i}(x)
&=&
 -\varepsilon_{ij}\,\frac{x_j}{r^2}\
\left(1-f_3(r)\right),\;
\nonumber\\[4mm]
A_{i}(x)
&=&
- \varepsilon_{ij}\,\frac{x_j}{r^2}\
\left(1-f(r)\right)\,,
\label{znstr}
\eeqn
where $i=1,2$ labels coordinates in the plane orthogonal to the string
axis and $r$ and $\alpha$ are the polar coordinates in this plane. The profile
functions $\phi_1(r)$ and  $\phi_2(r)$ determine the profiles of
the scalar fields,
while $f_{3}(r)$ and $f(r)$ determine the SU($2$) and U(1)
gauge fields of the
string solution, respectively. These functions satisfy the following
first-order equations \cite{ABEKY}:
\beqn
&&
r\frac{d}{{d}r}\,\phi_1 (r)- \frac12\left( f(r)
+  f_3(r) \right)\phi_1 (r) = 0\, ,
\nonumber\\[4mm]
&&
r\frac{d}{{ d}r}\,\phi_2 (r)- \frac12\left(f(r)
-  f_3(r)\right)\phi_2 (r) = 0\, ,
\nonumber\\[4mm]
&&
-\frac1r\,\frac{ d}{{ d}r} f(r)+\frac{g^2_1}{2}\,
\left[\left(\phi_1(r)\right)^2 +\left(\phi_2(r)\right)^2-2\xi\right] =
0\, ,
\nonumber\\[4mm]
&&
-\frac1r\,\frac{d}{{ d}r} f_3(r)+\frac{g^2_2}{2}\,
\left[\left(\phi_1(r)\right)^2 -\left(\phi_2(r)\right)^2\right]  = 0
\, .
\label{foe}
\eeqn
The boundary conditions
for the profile functions in these equations are
\beqn
&&
f_3(0) = 1\, ,\qquad f(0)=1\, ;
\nonumber\\[4mm]
&&
f_3(\infty)=0\, , \qquad   f(\infty) = 0
\label{fbc}
\eeqn
for the gauge fields, while the boundary conditions for  the
squark fields are
\beqn
\phi_1 (\infty)=\sqrt{\xi}\,,\qquad   \phi_2 (\infty)=\sqrt{\xi}\,,
\qquad \phi_1 (0)=0\, .
\label{phibc}
\eeqn
Note that since the field $ \phi_2 $ does not wind, it need not vanish
at the origin, and, in fact, it does not. Numerical solutions of the
Bogomol'nyi equations (\ref{foe}) for the $Z_2$ strings were
found in Ref.~\cite{ABEKY}.

The tension of the elementary $Z_2$ string is
\beq
T=2\pi\,\xi\, ,
\label{ten}
\eeq
to be compared with  the tension of the Abelian Abrikosov--Nielsen--Olesen (ANO)
string \cite{ANO},
\beq
T_{\rm ANO}=4\pi\,\xi
\label{tenANO}
\eeq
in our normalization.

Making the elementary $Z_2$ strings 
``wind" in SU(2) makes it {\em bona fide} non-Abelian.
This means that, besides trivial translational
moduli, the string acquires SU(2)/U(1) moduli. Indeed, while the ``flat"
vacuum (\ref{qvev}) is SU($2$)$_{C+F}$ symmetric, the solution (\ref{znstr})
breaks this symmetry down to U(1) which gives rise to a family of degenerate
solutions.

To obtain the above family from the $Z_2$ string
(\ref{znstr}) we act on it by diagonal color-flavor rotations preserving
the vacuum (\ref{qvev}). To this end
it is convenient to pass to the singular gauge where the scalar fields have
no winding at infinity, while the string flux comes from the vicinity of
the origin. In this gauge we have
\begin{eqnarray}
&&
q
=
U \left(
\begin{array}{cc}
\phi_1(r) & 0  \\[2mm]
0 &  \phi_2(r)
\end{array}\right)U^{-1}\,=\, \frac12 (\phi_1+\phi_2)
+\frac{\tau^a}{2}S^a (\phi_1-\phi_2) ,
\nonumber \\[4mm]
&&
A^a_{i}(x)
=
S^a \,\varepsilon_{ij}\, \frac{x_j}{r^2}\,
f_3(r)\, ,\qquad
A_{i}(x)
= \varepsilon_{ij} \, \frac{x_j}{r^2}\,
f(r)\, ,
\label{sna}
\end{eqnarray}
where $U$ is a matrix $\in {\rm SU}(2)$ and
$ S^a$ is a moduli vector
defined as
\beq
S^a\tau^a=U \tau^3 U^{-1},\;\;a=1,2,3,
\label{n}
\eeq
and subject to the constraint
\beq
{\vec S}^{\,2} = 1\,.
\label{ensq}
\eeq
At $S=\{0,0, 1\}$ we get the field configuration quoted
in Eq.~(\ref{znstr}).

As soon as the SU(2)$_{C+F}$ group is broken by the string solution
(\ref{znstr}) down to U(1), the effective two-dimensional theory on the string
which describes the internal dynamics of the orientational moduli $S^a$ is 
the $O(3) = CP(1)$ model ($CP(N-1)$ in the general case of the SU$(N)\times$U(1) gauge group in the bulk theory)
\cite{HT1,ABEKY,SYmon,HT2,GSY05}. The bosonic action has the form
(for derivation see the review paper \cite{SYrev})
\beq
S^{(1+1)}= \frac{ \beta}{2}\,   \int d t\, dz \,
\left(\pt_k\,  S^a\right)^2\,,
\label{o3}
\eeq
where the coupling constant $\beta$ is given by a normalizing integral
\beq
\beta=
\frac{2\pi}{g_2^2}\,  \int_0^{\infty}
dr\left\{-\frac{d}{dr}f_3
+\left(\frac{2}{r}\, f_3^2+\frac{d}{dr}f_3\right)\,\frac{\phi_1^2}{\phi_2^2}
\right\}.
\label{beta}
\eeq
Using the first-order equations for the string profile functions (\ref{foe})
one can see that
the integral  here reduces to a total derivative and is given
by the flux of the string  determined by $f_3(0)=1$. Thus
\beq
\beta= \frac{2\pi}{g_2^2}\,.
\label{betag}
\eeq
The two-dimensional coupling constant is determined by the
four-dimensional non-Abelian coupling.

The above relation between the four-dimensional and two-dimensional coupling
constants (\ref{betag}) is obtained  at the classical level. In quantum theory
both couplings run. In particular, the $CP(1)$ model is asymptotically free
\cite{Po3} and develops its own scale $\Lambda_{CP(1)}$. Its relation to the 
parameters of the bulk theory (\ref{model}) is given by 
\beq
\Lambda_{CP(1)}=\frac{\Lambda_{{\mathcal N}=1}^2}{m_W}\,,
\label{cpscale2}
\eeq
see Ref.~\cite{SYnone}.

\subsection{Fermion zero modes: \boldmath\ntwo limit}
\label{fermzeromodes}

Let us start from the \ntwo theory (\ref{model}) with no deformation 
superpotential. Our string solution is 1/2 BPS-saturated. This means that four
supercharges, out of eight of the four-dimensional theory
(\ref{model}), act trivially on the string solution
(\ref{sna}). The remaining four supercharges generate four fermion zero
modes which we call supertranslational modes because they are
superpartners to  two translational zero modes. The corresponding four fermionic moduli
are superpartners to the coordinates $x_0$ and
$y_0$ of the string center. The  supertranslational
fermion zero modes were found in Ref.~\cite{VY} for the Abelian ANO string.
Below we generalize this construction to the case of the non-Abelian string.

The  fermionic part of the action  of  the model (\ref{model})
is
\beqn
S_{\rm ferm}
&=&
\int d^4 x\left\{
\frac{i}{g_2^2}\bar{\lambda}_f^a \bar{D}\hspace{-0.65em}/\lambda^{af}+
\frac{i}{g_1^2}\bar{\lambda}_f \bar{\pt}\hspace{-0.65em}/\lambda^{f}
+ {\rm Tr}\left[\bar{\psi} i\bar\nabla\hspace{-0.65em}/ \psi\right]
+ {\rm Tr}\left[\tilde{\psi} i\nabla\hspace{-0.65em}/ \bar{\tilde{\psi}}
\right]\right.
\nonumber\\[3mm]
&+&
\frac{i}{\sqrt{2}}\,{\rm Tr}\left[ \bar{q}_f(\lambda^f\psi)+
(\tilde{\psi}\lambda_f)q^f +(\bar{\psi}\bar{\lambda}_f)q^f+
\bar{q}^f(\bar{\lambda}_f\bar{\tilde{\psi}})\right]
\nonumber\\[3mm]
&+&
\frac{i}{\sqrt{2}}\,{\rm Tr}\left[ \bar{q}_f\tau^a(\lambda^{af}\psi)+
(\tilde{\psi}\lambda_f^a)\tau^aq^f +(\bar{\psi}\bar{\lambda}_f^a)\tau^aq^f+
\bar{q}^f\tau^a(\bar{\lambda}^{a}_f\bar{\tilde{\psi}})\right]
\nonumber\\[3mm]
&+&
\frac{i}{\sqrt{2}}\,{\rm Tr}\left[\tilde{\psi}\left(a+a^a\tau^a\right)\psi
\right]
+\frac{i}{\sqrt{2}}\,{\rm Tr}\left[\bar{\psi}\left(a+a^a\tau^a\right)
\bar{\tilde{\psi}}\right]
\nonumber\\[3mm]
&-&
\left. \frac{\mu}2 (\lambda^2)^2
-\frac{\mu}2 (\lambda^{a2})^2
\right\}\,,
\label{fermact}
\eeqn
where the  matrix color-flavor notation is used for the
matter fermions $(\psi^{\alpha})^{kA}$ and $(\tilde{\psi}^{\alpha})_{Ak}$.
The traces are performed
over the color-flavor indices. Contraction of the spinor indices is assumed
inside all parentheses, for instance,
$(\lambda\psi)\equiv \lambda_{\alpha}\psi^{\alpha}\,,$ see Appendix A.
We write the squark fields in (\ref{fermact}) as doublets of the SU(2)$_{R}$
group which is present in \ntwo theory, $q^f=(q,\bar{\tilde{q}})$. Here
$f=1,2$ is the SU(2)$_R$ index
which labels two supersymmetries of the bulk theory in the \ntwo limit.
Moreover,
$\lambda^{\alpha f}$ and $(\lambda^{\alpha f})^a$ stand for
the gauginos of the U(1) and SU(2) groups, respectively.
Note that the last two  terms are \none deformations
in the fermion sector of the theory induced by the breaking parameter $\mu$.
They involve only $f=2$ components of $\lambda$'s explicitly breaking
the SU(2)$_{R}$ invariance.

Now, we put $\mu=0$ (consideration of $\mu\neq 0$ will be carried out
in Sect.~\ref{N1zeromodes}) and
apply supersymmetry transformations to generate four supertranslational modes
of the non-Abelian string  in the \ntwo limit.
The supertransformations in our bulk theory have the form
\begin{eqnarray}
\delta\lambda^{f\alpha}
&=&
\frac12(\sigma_\mu\bar{\sigma}_\nu\epsilon^f)^\alpha
F_{\mu\nu}+\epsilon^{\alpha p}D^m(\tau^m)^f_p\ +\dots,
\nonumber\\[3mm]
\delta\lambda^{af\alpha}
&=&
\frac12(\sigma_\mu\bar{\sigma}_\nu\epsilon^f)^\alpha
F^a_{\mu\nu}+\epsilon^{\alpha p}D^{am}(\tau^m)^f_p\ +\dots,
\nonumber\\[3mm]
\delta\bar{\tilde\psi}_{\dot{\alpha}}^{kA}
&=&
i\sqrt2\
\bar\nabla\hspace{-0.65em}/_{\dot{\alpha}\alpha}q_f^{kA}\epsilon^{\alpha
f}\ +\cdots,
\nonumber\\[4mm]
\delta\bar\psi_{\dot{\alpha}Ak}
&=&
 i\sqrt2\
\bar\nabla\hspace{-0.65em}/_{\dot{\alpha}\alpha}\bar
q_{fAk}\epsilon^{\alpha f}\ +\cdots.
\label{transf}
\end{eqnarray}
Here the parameters of SUSY transformations
 are denoted as  $\epsilon^{\alpha f}$.
Furthermore, the $D$ terms in Eq.~(\ref{transf}) are
\beq
D^1+ iD^2 =0
\, ,\qquad 
D^3 =-i\, \frac{g_1^2}{2}\, \left( {\rm Tr}\, |q|^2-2\xi\right)
\label{pdterm}
\eeq
for the U(1) field, and
\beq
D^{a1}+iD^{a2}=0
\,,
\qquad
D^{a3}=-i\, \frac{g_2^2}{2}\,{\rm Tr}\, \left(\bar{q}\tau^a q\right)
\label{vdterm}
\eeq
for the SU(2) field. The dots in (\ref{transf})  stand for terms involving
the adjoint scalar fields which vanish on the string solution
(at $m_1=m_2$) because the
adjoint fields are given by their vacuum expectation values
(\ref{avev}).

In Ref.~\cite{VY} it was shown that the four supercharges associated with
the parameters $\epsilon^{12}$ and $\epsilon^{21}$
act trivially on the BPS string in the theory with the
Fayet--Iliopoulos $D$ term. The same is true for the non-Abelian string
solution (\ref{sna}). Applying supertransformations (\ref{transf}) with the
parameters $\epsilon^{11}$ and $\epsilon^{22}$ to (\ref{sna})
we generate the following supertranslational zero modes:
\beqn
\bar{\psi}_{Ak\dot{2}}
& = &
-\,\frac1{\sqrt{2}}\,\frac{x_1+ix_2}{r^2}\left\{\left[(f+f_3)\phi_1+(f-f_3)\phi_2\right]
\right.
\nonumber\\[4mm]
&+&
\left.
\tau^a S^a\,\left[(f+f_3)\phi_1-(f-f_3)\phi_2\right]\right\}\zeta_L
\, ,
\nonumber\\[4mm]
\bar{\tilde{\psi}}^{kA}_{\dot{1}}
& = &
\,\frac1{\sqrt{2}}\,\frac{x_1-ix_2}{r^2}\left\{\left[(f+f_3)\phi_1+(f-f_3)\phi_2\right]
\right.
\nonumber\\[4mm]
&+&
\left.
\tau^a S^a\,\left[(f+f_3)\phi_1-(f-f_3)\phi_2\right]\right\}\zeta_R
\, ,
\nonumber\\[4mm]
\bar{\psi}_{Ak\dot{1}}
& = &
0\, , \qquad
\bar{\tilde{\psi}}^{kA}_{\dot{2}}= 0\, ,
\nonumber\\[4mm]
\lambda^{a22}
& = &
ig^2_2\,(\phi_1^2-\phi_2^2)\,S^a\,\zeta_R
\, ,
\nonumber\\[4mm]
\lambda^{a11}
& = &
-ig^2_2\,(\phi_1^2-\phi_2^2)\,S^a\,\zeta_L
\,,
\nonumber\\[4mm]
\lambda^{a12}
& = & 0
  \, ,\qquad  \lambda^{a21}= 0\,,
\nonumber\\[4mm]
\lambda^{22}
& = &
ig^2_1\,(\phi_1^2+\phi_2^2-2\xi)\,\zeta_R
\, ,
\nonumber\\[4mm]
\lambda^{11}
& = &
-ig^2_1\,(\phi_1^2+\phi_2^2-2\xi)\,\zeta_L
\,,
\nonumber\\[4mm]
\lambda^{12}
& = & 0
  \, ,\qquad  \lambda^{21}= 0\,,
\label{tzmodes}
\eeqn
where the dependence on $x_i$ is encoded in the string profile
functions, see Eq.~(\ref{sna}), while the Grassmann parameters $\zeta_L$ and
$\zeta_R$ are related to the SUSY transformation parameters,
\beq
\delta \zeta_L=\epsilon^{11},\qquad \delta \zeta_R=\epsilon^{22}.
\label{zeta}
\eeq
These parameters become superpartners of the  string center coordinates
$x_i$ ($i=1,2$) in the effective theory on the string worldsheet.

Besides four supertranslational modes the non-Abelian string has four
superorientational modes. They were calculated in \cite{SYmon} using supersymmetry
transformations (\ref{transf}) with the parameters $\epsilon^{12}$ and $\epsilon^{21}$.
They have the following form:
\beqn
\bar{\psi}_{Ak\dot{2}}
& = &
\left(\frac{\tau^a}{2}\right)_{Ak}
\frac1{2\phi_2}(\phi_1^2-\phi_2^2)
\left[
\chi_L^a
+i\varepsilon^{abc}\, S^b\, \chi^c_L\,
\right]\, ,
\nonumber\\[3mm]
\bar{\tilde{\psi}}^{kA}_{\dot{1}}
& = &
\left(\frac{\tau^a}{2}\right)^{kA}
\frac1{2\phi_2}(\phi_1^2-\phi_2^2)
\left[
\chi_R^a
-i\varepsilon^{abc}\, S^b\, \chi^c_R\,
\right]\, ,
\nonumber\\[5mm]
\bar{\psi}_{Ak\dot{1}}
& = &
0\, , \qquad
\bar{\tilde{\psi}}^{kA}_{\dot{2}}= 0\, ,
\nonumber\\[4mm]
\lambda^{a22}
& = &
\frac{i}{\sqrt{2}}\frac{x_1+ix_2}{r^2}
f_3\frac{\phi_1}{\phi_2}
\left[
\chi^a_R
-i\varepsilon^{abc}\, S^b\, \chi^c_R
\right]\, ,
\nonumber\\[4mm]
\lambda^{a11}
& = &
\frac{i}{\sqrt{2}}\frac{x_1-ix_2}{r^2}
f_3\frac{\phi_1}{\phi_2}
\left[
\chi^a_L
+i\varepsilon^{abc}\, S^b\, \chi^c_L
 \right]\,,
\nonumber\\[4mm]
\lambda^{a12}
& = & 0
  \, ,\qquad  \lambda^{a21}= 0\,,
\label{ozmodes}
\eeqn
where $\chi_L^a$ and $\chi_R^a$ are real Grassmann parameters, subject to constraints
\beq
S^a \chi_L^a=0\,,\qquad S^a \chi_R^a=0\, .
\label{constraint}
\eeq

We can directly verify that the zero modes (\ref{tzmodes}) and (\ref{ozmodes})
satisfy the Dirac equations of motion. From the fermion action 
of the model (\ref{fermact}) we get
the relevant Dirac equations for $\lambda^a$,
\beqn
&&\frac{i}{g_1^2} \bar{D}\hspace{-0.65em}/\lambda^{f}
+\frac{i}{\sqrt{2}}\,{\rm Tr}\left(
\bar{\psi}q^f+
\bar{q}^f\bar{\tilde{\psi}}\right)-
\mu\delta^f_2\bar{\lambda}_2=0\,,
\nonumber\\[3mm]
&&\frac{i}{g_2^2} \bar{D}\hspace{-0.65em}/\lambda^{af}
+\frac{i}{\sqrt{2}}\,{\rm Tr}\left(
\bar{\psi}\tau^a q^f+
\bar{q}^f\tau^a\bar{\tilde{\psi}}\right)-
\mu\delta^f_2\bar{\lambda}_2^a=0\,,
\label{dirac1}
\eeqn
while  for the matter fermions
\beqn
&&
i\nabla\hspace{-0.65em}/ \bar{\psi}+\frac{i}{\sqrt{2}}\left[\bar{q}_f\lambda^f
-(\tau^a\bar{q}_f)\lambda^{af}+
(a-a^a\tau^a)\tilde{\psi}\right]=0\, ,
\nonumber\\[3mm]
&&
i\nabla\hspace{-0.65em}/ \bar{\tilde{\psi}}+
\frac{i}{\sqrt{2}}\left[\lambda_f q^f
+\lambda^{a}_f(\tau^aq^f)+
(a+a^a\tau^a)\psi\right]=0\, .
\label{dirac2}
\eeqn
Now  we
substitute the supertranslational and superorientational fermion zero modes 
(\ref{tzmodes}) and (\ref{ozmodes}) into these
equations in the limit $\mu=0$. After some algebra we managed to check
that  they do  satisfy the Dirac equations (\ref{dirac1}) and
(\ref{dirac2})
provided the first-order equations for the string profile functions (\ref{foe})
are fulfilled (this check for superorientational modes was done in \cite{SYnone}).

\subsection{ \boldmath{$CP(1)\times C$} model on the string
worldsheet: direct\\ calculation in the \ntwo limit}
\label{2,2}

The zero modes (\ref{tzmodes}) and (\ref{ozmodes}) generate
the fermion part of  the \ntwot model with the target space $CP(1)\times C$.
This statement was checked in \cite{SYnone}. To perform the check we assume, as usual,
that the fermion collective coordinates  $\zeta_{L,R}$ and $\chi^a_{L,R}$ 
have an adiabatic  dependence on the worldsheet coordinates $x_k$ ($k=0,3$). 
Substituting (\ref{tzmodes}) and (\ref{ozmodes}) in  the fermion kinetic terms
in the bulk  theory (\ref{fermact}),
and taking into account the derivatives of   $\zeta_{L,R}$ and $\chi^a_{L,R}$ with
respect to the worldsheet coordinates  we arrive at
\beqn
&& 
S_{1+1}
=
 \int d t d z \left\{2\pi\xi\left[\frac12
\left(\pt_k x_{0i}
\right)^2+
\frac12 \, \bar{\zeta}_R \, i
\left(\pt_0-i\pt_3
\right)\, \zeta_R
\right.
+
\frac12 \, \bar{\zeta}_L \, i
\left(\pt_0+i\pt_3
\right)\, \zeta_L
\right]
\nonumber\\[3mm]
&&+
\beta \left[\frac12 (\pt_k S^a)^2+
\frac12 \, \chi^a_R \, i(\pt_0-i\pt_3)\, \chi^a_R
+ 
\left.
\frac12 \, \chi^a_L \, i(\pt_0+i\pt_3)\, \chi^a_L
-\frac12 (\chi^a_R\chi^a_L)^2
\right]
\right\},
\nonumber\\
\label{ntwocp}
\eeqn
where $x_{0i}$ ($i=1,2$) denote the coordinates of the string center
in $(1,2)$-plane; the value of  $\beta$ is determined by the same integral (\ref{beta}) 
as in the the bosonic kinetic term, see Eq.~(\ref{o3}). The first line corresponds to
the $C$ part of the target space, while the second line to the
$CP(1)$ part. 
The model specified by the second line in Eq.~(\ref{ntwocp})
(plus the constraint (\ref{constraint}))
is also known as supersymmetric O(3) sigma model \cite{ewi,pdive}.
In the \ntwo limit all three fields,
$x_0$,   $\zeta_L$ and $\zeta_R$ are sterile. Deformations to be discussed below
will leave $x_0$ and $\zeta_L$ sterile, while at the same time
will couple $\zeta_R$ with the $CP(1)$ sector.

In fact, our derivation gives only the quadratic terms in the fermion fields.
The four-fermion term is not accessible in this approximation.
The worldsheet \ntwot supersymmetry was used in \cite{SYnone} to reconstruct the
four-fermion interactions inherent to $CP(1)$. The SUSY transformations
in the $CP(1)$ model have the form (see e.g. \cite{NSVZsigma})
\beqn
\delta \chi^a_R
&=&
i\sqrt{2} \left( \pt_0 +i\pt_3\right) S^a \, \varepsilon_2
+\sqrt{2}\varepsilon_1\,S^a(\chi^a_R\chi^a_L)\,,
\nonumber\\[3mm]
\delta \chi^a_L
&=&
i\sqrt{2} \left( \pt_0 -i\pt_3\right)  S^a \, \varepsilon_1
-\sqrt{2}\varepsilon_2\,S^a(\chi^a_R\chi^a_L)\, ,
\nonumber\\[3mm]
\delta S^a
&=&
\sqrt{2}(\varepsilon_1\chi^a_L+\varepsilon_2\chi^a_L)\,,
\label{susy2d}
\eeqn
where  $\varepsilon_{1,2}$ are two parameters of extended \ntwot transformations
(for simplicity we  restrict ourselves to the real parts of 
these transformations). Imposing this supersymmetry
fixes the four-fermion term in (\ref{ntwocp}).

$CP(N-1)$ 
model (which is a part of string worldsheet theory) can be nicely rewritten in
terms of two-dimensional U(1) gauge theory of $N$ complex fields $n^l$
in the strong coupling ($e^2\to\infty$) limit \cite{W93}. This is the so-called
gauged formulation in which the bosonic part of the action takes the form
\beqn
S_{CP(1)\,{\rm bos}}
& =&
\int d^2 x \left\{
 |\nabla_{k} n^{l}|^2 +\frac1{4e^2}F^2_{kl} + \frac1{e^2}
|\pt_k\sigma|^2+\frac1{2e^2}D^2
\right.
\nonumber\\[3mm]
 &+&    2|\sigma|^2 |n^{l}|^2 + iD (|n^{l}|^2-2\beta)
\Big\}\,,
\label{cpg}
\eeqn
where 
$$
\nabla_k= \partial_k - i A_k \,,\quad F_{kl} = \partial_k A_l -
\partial_l A_k\,,
$$
while $\sigma$ is a complex scalar field and $D$ is the $D$-component of the gauge multiplet. Eliminating the $D$-component leads to the constraint
\beq
|n^l|^2=2\beta\,.
\label{nconstraint}
\eeq
Moreover, in the  limit $e^2\to\infty$
the gauge field $A_k$  and its \ntwo bosonic superpartner $\sigma$ become
auxiliary (their kinetic terms vanish) and can be eliminated by virtue of the equations of motion (the fermion fields are ignored so far),
\beq
A_k =-\frac{i}{4\beta}\, \bar{n}_l \stackrel{\leftrightarrow}
{\partial_k} n^l \,, \qquad \sigma = 0\,.
\label{aandsigma}
\eeq
With $2N$ complex fields $n^l$, one real constraint (\ref{nconstraint}) and one phase
``eaten'' by gauging  U(1), the model has $2N-1-1=2(N-1)$ real degrees of freedom.

At $N=2$ Eq.~(\ref{cpg}) is equivalent to the bosonic action of O(3) sigma model
(\ref{o3}). The relation between the variables $S^a$ and $n^l$ is
\beq
S^a=\frac1{2\beta}\,\bar{n}\tau^a n\, .
\label{Sn}
\eeq

The fermionic part of the $CP(N-1)$ model action written in the gauged formulation
has the form
\beqn
S_{CP(1)\,{\rm ferm}}
& =&
\int d^2 x \left\{
 \bar{\xi}_{lR}\,i(\nabla_{0}-i\nabla_3) \xi^{l}_R
+ \bar{\xi}_{lL}\,i(\nabla_{0}+i\nabla_3) \xi^{l}_L
 \right.
\nonumber\\[3mm]
 &+& 
\frac1{e^2}\,\bar{\lambda}_{R}\,i(\nabla_{0}-i\nabla_3) \lambda_R
+\frac1{e^2}\,\bar{\lambda}_{L}\,i(\nabla_{0}+i\nabla_3) \lambda_L
+i\sqrt{2}\,\sigma\,\bar{\xi}_{lR}\xi^l_L
\nonumber\\[3mm]
 &+& 
\left.
 i\sqrt{2}\,\bar{n}_l\,(\lambda_R\xi^l_L-
\lambda_L\xi^l_R) + {\rm c.c.}  
\right\}\,,
\label{cpgf}
\eeqn
where the fields $\xi^l_{L,R}$ are fermion superpartners of $n^l$ while $\lambda_{L,R}$
belong to the gauge multiplet. In the limit $e^2\to\infty$ the fields
$\lambda_{L,R}$ become auxiliary implying the following constraints:
\beq
\bar{n}^l \xi^l_L=0, \qquad \bar{n}^l \xi^l_R=0\,.
\label{nxiconstraint}
\eeq
With the fermions switched on Eq.~(\ref{aandsigma}) must be replaced by
\beqn
A_0+iA_3 
& = &
-\frac{i}{4\beta}\, \bar{n}_l \left(\stackrel{\leftrightarrow}{\partial_0}
+i\stackrel{\leftrightarrow}{\partial_3}\right) 
n^l -\frac{1}{2\beta}\,\bar{\xi}_R\xi_R\,,
\nonumber\\[3mm]
A_0-iA_3 
& = &
-\frac{i}{4\beta}\, \bar{n}_l \left(\stackrel{\leftrightarrow}{\partial_0}
-i\stackrel{\leftrightarrow}{\partial_3}\right) 
n^l -\frac{1}{2\beta}\,\bar{\xi}_L\xi_L\,,
\nonumber\\[3mm]
\sigma
& = &
-\frac{i}{2\sqrt{2}\beta}\;\bar{\xi}_{lL}\xi^l_R\,.
\label{sigma} 
\eeqn
The extra $\xi$ terms in $A_k$ and $\sigma$
are responsible for the four-fermion part of the Lagrangian in the gauged formulation.

At $N=2$, the theory  specified in
(\ref{cpg}), (\ref{cpgf}) and (\ref{sigma}) is equivalent to the O(3) sigma model
(\ref{ntwocp}). The relation between the complex fermions
$\xi^l$ of the gauged formulation and real fermions $\chi^a$ 
of the O(3) sigma model is  
\beq
\chi^a_{L,R}=\frac1{2\beta}\,\left(\bar{n}\,\tau^a\xi_{L,R}+
\bar{\xi}_{L,R}\,\tau^a n\right)\,.
\label{chixi}
\eeq

\section{
 Digression: Edalati--Tong's suggestion}
\label{0,2}
\setcounter{equation}{0}

The previous part of the paper, along with new results for the fermion zero modes,
contained many elements of a review nature.
Now we are finally ready to venture into uncharted waters
which will bring us, eventually, to `` heterotic $CP(1)$."

Let us break \ntwo supersymmetry of the bulk theory by switching on the deformation
superpotential of the type (\ref{msuperpotbr}) or (\ref{defsup}).
In both cases the field $\tilde{q}$ has vanishing
VEV, and the string solutions remains BPS-saturated \cite{SYnone,Edalati}.

The case of the adjoint mass deformation (\ref{msuperpotbr}) was considered 
in detail in \cite{SYnone}.
With four supercharges of the  deformed \none bulk theory normally
the 1/2 BPS-saturated string solution (\ref{sna}) will preserve
only two supercharges on the string worldsheet. However, the number of the fermion zero
modes on the string does not change when we break \ntwo by virtue of the
superpotential (\ref{msuperpotbr}). This number is fixed by the index theorem
obtained in \cite{GSYmmodel}.
Thus, the number of (classically) massless fermion fields on the worldsheet 
does not change. It is well-known that the sigma model with the $CP(N-1)$ target
space, when supersymmetrized, automatically yields \ntwot sigma model; one cannot
get \ntwoo$\!$. Therefore, in Ref.~\cite{SYnone} it was concluded that the worldsheet theory has an ``accidental'' SUSY enhancement.

On the other hand, in the recent publication \cite{Edalati} it was 
pointed out that the target space in the problem at hand is $CP(N-1)\times C$
rather than $CP(N-1)$. Edalati and Tong suggested that
the superorientational zero modes can mix with supertranslational ones.
In fact, even earlier it was noted  \cite{rit} that such a mixing, if it takes place, could occur only through a modification of the constraint (\ref{nxiconstraint}),
$$
\bar{n}^l \xi^l_R \propto \bar{\zeta}_R
$$
in the case of the monomial deformation (\ref{msuperpotbr}). Moreover, Edalati and Tong
explicitly constructed an \ntwoo supergeneralization of the sigma model with
the target space $CP(N-1)\times C$.
In their construction the \ntwot model (\ref{cpg}), (\ref{cpgf})
is supplemented by the term
\beq
\delta S_{1+1}=\int d^2 x\,
2\beta\,  \left\{4\left|\frac{\pt {\mathcal W}_{1+1}}{\pt \sigma}
\right|^2 + im_W\,\bar{\lambda}_{L}\,\frac{\pt^2 {\mathcal W}_{1+1}}{\pt \sigma^2}
\,\zeta_R\right\}
\label{etong}
\eeq
breaking \ntwot down to \ntwoo$\!$. Here ${\mathcal W}_{1+1}$ is a two-dimensional deformation superpotential while $m_W$ is the mass of the SU(2) gauge boson (\ref{wmass}). Integrating out the axillary field $\lambda$ now leads us to
\beq
\bar{n}^l \,\xi^l_L=0, \qquad \bar{\xi}_{lR}\, n^l=\sqrt{2}\beta\, m_W\,
\frac{\pt^2 {\mathcal W}_{1+1}}{\pt \sigma^2}
\,\zeta_R\, .
\label{modnxiconstraint}
\eeq
The left-handed fermion sector remains intact, while the right-handed fermion sector
changes. The constraint (\ref{nxiconstraint}) is modified: the right-handed
fermion $\zeta_R$ from the translational sector no longer decouples from 
the orientational one. The left-handed fermion $\zeta_L$ as well as $x_0$
remain free fields and can be omitted in what follows. We suggest a concise name for the model obtained in this way: `` heterotic $CP(1)$."
The general structure of the deformation
(\ref{etong}) is dictated by \ntwoo supersymmetry.

Our analysis fully confirms the above statements. However, this is not the end of the story: Edalati and Tong suggested, additionally, that the bulk and worldsheet
deformation superpotentials coincide, 
\beq
{\mathcal W}_{1+1} \sim {\mathcal W}_{3+1}\,,
\label{etconjecture}
\eeq
implying that this coincidence takes place for all superpotentials
of the type (\ref{msuperpotbr}) and (\ref{defsup}). The analysis to be presented below
shows that Eq.~(\ref{etconjecture}) is valid only at small $\mu$,
to the leading order in $\mu$. To this order the worldsheet theory deformation is determined
essentially by the critical points of the superpotential. At finite or large $\mu$
the worldsheet deformation superpotential is not given by the simple formula (\ref{etconjecture}).
In particular, for the deformation (\ref{defsup}) it becomes nonpolynomial.
In Sects.~\ref{02}--\ref{crossterm}
we study the bulk deformation (\ref{msuperpotbr}) and then in Sect. \ref{twistedmass} turn to the bulk deformation (\ref{defsup}).
In the remainder of this paper we will derive the string worldsheet theory
starting from the \none bulk theory with the
deformation superpotentials (\ref{msuperpotbr}) or (\ref{defsup}).

Let us first dwell on (\ref{msuperpotbr}) which implies ${\mathcal W}_{1+1}\propto\sigma^2$. 
The gauged formulation exploited by Edalati and Tong is convenient for establishing a general
structure of the two-dimensional \ntwoo sigma model. However, it is inconvenient
if one's goal is a direct derivation of this worldsheet model form
the bulk theory. The reason is rather obvious: in the gauged formulation both the bosonic part and two-fermion terms in the worldsheet Lagrangian are the same for \ntwot and \ntwoo$\!$. Therefore, to detect the difference one  has to deal with four-fermion 
terms whose extraction from the bulk theory is technically very
difficult. At the same time, as we will see shortly, in the O(3) formulation with 
the undeformed constraints (\ref{constraint})
the difference between \ntwot and \ntwoo shows up in two-fermion terms
which are readily calculable from the bulk theory given our knowledge of the fermion zero modes.

In the next section we will prove the above statement by
deriving \ntwoo super\-generalization of the O(3) sigma model. To this end we will need
to redefine the field $\xi_R$ by introducing a linear combination 
of $\xi_R$ and $\zeta_R$ such that for the new field
$
\xi_R'  =\xi_R -{\rm const}\,n\,\bar{\zeta}_R
$
we have the conventional constraint $\bar n \, \xi_R' =0$. Then we can use Eqs.~(\ref{Sn})  and (\ref{chixi}) (with $\xi$ replaced by $\xi' $) to pass to the O(3) formulation.
The constraint $\vec S \vec\chi_{L,R}=0$
will be satisfied automatically. In Sect.~\ref{02} we will present the result of this construction
assuming that the deformation superpotential is that of the mass term.

\section{Heterotic \boldmath{$CP(1)$}}
\label{02}
\setcounter{equation}{0}

Here we will derive the \ntwoo supergeneralization of the O(3) sigma model 
following the program outlined above.
We will assume for the time being that the deformation potential is monomial, see 
Eq.~(\ref{msuperpotbr}), or equivalently,\footnote{The relation between 
$\delta$ and $\mu$ will be established in Sect.~\ref{crossterm}.
According to the Edalati--Tong conjecture $\delta\sim \mu$.
We will see that at small $\mu$ proportionality between
$\mu$ and $\delta$ does take place. At large $\mu$, as  we will see below,
$\delta$ grows as $\ln \mu$.}
\beq
{\mathcal W}_{1+1}= \frac{\delta}{2}\; \sigma^2\,,
\label{w11}
\eeq
where $\delta$ is a constant (see below and Appendix D).
Performing a rather straightforward algebraic analysis
we get
\beqn
S_{1+1}
&=&
 \int d^2 x \left\{2\pi\xi
 \,
 \left[\frac12
\left(\pt_k \vec{x}_{0} \right)^2+
\frac12 \, \bar{\zeta}_L \, i\,\partial_R \, \zeta_L
+  
\frac12 \, \bar{\zeta}_R \, i\,\partial_L \, \zeta_R
\right]\right.
\nonumber\\[3mm]
&+& 
\beta \, \left[\frac12 \left(\pt_k S^a\right)^2+
\frac12 \, \chi^a_R \, i\,\partial_L \, \chi^a_R
+
\frac12 \, \chi^a_L \, i\,\partial_R\, \chi^a_L
-\frac{c^2}{2} (\chi^a_R\chi^a_L)^2
\right.
\nonumber\\[4mm]
&+&
c\,\chi_R^a\left(\rule{0mm}{5mm}
i\,\partial_L\,S^a\,(\kappa\,\zeta_R +\bar{\kappa}\,\bar{\zeta}_R)+i\varepsilon^{abc} S^b
i\,\partial_L\,S^c\,(\kappa\,\zeta_R-\bar{\kappa}\,\bar{\zeta}_R)\right)
\nonumber\\[4mm]
&+&
\left.
\left.
2\,|\kappa|^2\,c^2\, \bar{\zeta}_R\zeta_R \,i\varepsilon^{abc}S^a\chi_L^b\chi_L^c
\rule{0mm}{5mm}\right]
\right\},
\label{02o3}
\eeqn
where 
the vector $\vec{x}$ parametrizes the position of the flux tube center  in the
perpendicular plane,
\beq
\partial_L\equiv \pt_0-i\pt_3\,,\quad \partial_R\equiv \pt_0+ i\pt_3\,,
\eeq
and
\beqn
\rule{0mm}{9mm}
c^2&=&\frac{1}{1+|\alpha|^2}\,,
\nonumber\\[3mm]
\alpha &\equiv& \frac{2\sqrt{2}\kappa}{m_W} = \frac{2\sqrt{2}\kappa}{g_2\sqrt{\xi}}\,.
\label{c}
\eeqn

The relation of the deformation parameter introduced here and the one in the gauged 
formulation of the theory (see Eqs.~(\ref{etong}) and  (\ref{w11})) is as follows:
\beq
\delta =\frac{\alpha}{\sqrt{1-|\alpha|^2}}\,.
\label{delta}
\eeq
To simplify reading of the paper we summarize  relations between different definitions of 
the deformation parameters of the worldsheet theory in Appendix D.

As was mentioned, the constraints $\rule{0mm}{5mm}S^a \chi_L^a=0$, $\, S^a \chi_R^a=0$
 for fermions
$\chi$ stay intact. This is achieved through shifting the field $\xi_R$,
\beq 
\bar{\xi}_R\to\bar{\xi}_R-  \delta\,\frac{m_W}{\sqrt{2}}\,\bar{n}\,{\zeta}_R\, ,
\label{shift}
\eeq
see Eq.~(\ref{modnxiconstraint}). As a result,  crucial bifermionic terms
of the type   $\chi_R\partial S\zeta_R$ 
appear in the third line of Eq.~(\ref{02o3}).

Generically the parameter $\kappa$ is complex. Its phase is determined by the bulk
deformation parameter $\mu$,
\beq
{\rm arg}\,\kappa={\rm arg}\,\mu .
\label{kappaphase}
\eeq
Later on,  for simplicity, we assume that the bulk parameter $\mu$ is real; therefore, 
$\kappa$ is real too.

The two-dimensional fields $\vec{x}_0$ and $\zeta_L$
forming a representation of \ntwoo super\-algebra are sterile;
they are decoupled from all other fields in the action
(\ref{02o3}). Although these fields are a part of the string worldsheet theory
they play no dynamical role. Therefore, in discussing dynamical aspects
of the heterotic model under consideration we
can (and will) safely omit them in what follows. Then the two-dimensional Lagrangian takes the form
\beqn
L_{1+1}
&=&
\beta \, \left[\frac12 \left(\pt_k S^a\right)^2+
\frac12 \, \chi^a_R \, i\,\partial_L \, \chi^a_R
+
\frac12 \, \chi^a_L \, i\,\partial_R\, \chi^a_L
-\frac{c^2}{2} (\chi^a_R\chi^a_L)^2
\right]
\nonumber\\[4mm]
&+&
\pi\xi \left(  \bar{\zeta}_R \, i\,\partial_L \, \zeta_R \right) +
\beta\Big[
c\,\kappa\,\chi_R^a\left(\rule{0mm}{5mm}
i\,\partial_L\,S^a\,(\zeta_R +\bar{\zeta}_R)+i\varepsilon^{abc} S^b
i\,\partial_L\,S^c\,(\zeta_R-\bar{\zeta}_R)\right)
\nonumber\\[4mm]
&+&
2\,\kappa^2\,c^2\, \bar{\zeta}_R\zeta_R \,i\varepsilon^{abc}S^a\chi_L^b\chi_L^c
\rule{0mm}{5mm}\Big] .
\label{02o3p}
\eeqn
The first line represents the conventional \ntwot $O(3)$ sigma model
(if we put $c=1$), 
while the second and the third lines give its \ntwoo deformation. The $\pi\xi$
normalizing factor in front of the kinetic term
$\bar{\zeta}_R \, i\,\partial_L \, \zeta_R $ is due to ``historical" reasons. The field
$\zeta_R $ used to be a superpartner of $\vec{x}_{0}$ in \ntwot$\!\!$.
As well-known, the normalization of the kinetic term of
$\vec{x}_{0}$ is given by the string tension. In Sect.~\ref{geom} we will switch to
the canonic normalization of the kinetic term
$\bar{\zeta}_R \, i\,\partial_L \, \zeta_R $.
The constant $\beta$ is related to the bulk constant $g_2^2$
and the conventionally normalized $O(3)$ model constant $g_0^2$ 
(see Sect.~\ref{geom}) as follows:
\beq
\beta =\frac{2\pi}{g_2^2} = \frac{1}{g_0^2}\,.
\label{relco}
\eeq
Let us ask ourselves: how many independent constants
characterize the heterotic $O(3)$ sigma model besides $g_0^2$?
At first sight, Eq.~(\ref{02o3p}) contains two constants, $\xi$ and $\kappa$.
In fact, there is only one extra constant. This is readily seen if one rescales the fields
$\zeta_R$ to make their kinetic tern canonically normalized.
Then one immediately sees that, besides $\beta=g_0^{-2}$, Eq.~(\ref{02o3p}) contains
a single additional constant, $\alpha$. This fact will be demonstrated again in 
Sect.~\ref{geom} where an alternative derivation of the heterotic deformation of the $CP(1)$ model is given. The relation between the (only) deformation parameter $\gamma$
introduced in Sect.~\ref{geom} and $\alpha$ is as follows:
\beq
\gamma^2 =\frac{1}{g_0^2}\, \frac{\alpha^2}{1+\alpha^2}=\beta \, \frac{\alpha^2}{1+\alpha^2}\,.
\label{gamal}
\eeq
We assume $\gamma$ to be real.

Concluding this section we present 
modified supertransformations.
The model (\ref{02o3}) is invariant under the following  \ntwoo SUSY
transformations:
\beqn
\delta \chi^a_R
&=&
\sqrt{2}\varepsilon_1\,S^a(\chi^b_R\chi^b_L)-\sqrt{2}\,{\kappa}\, {c}\, \varepsilon_1
\Big[ \chi_L^a\,(\zeta_R+\bar{\zeta}_R)\
\nonumber\\[3mm]
&+& i 
\, \varepsilon^{abc}\,  S^b \,\chi_L^c \left(
\zeta_R -\bar\zeta_R\right)\Big]\,,
\nonumber\\[3mm]
\delta \chi^a_L
&=&
i\sqrt{2} \left( \pt_0 -i\pt_3\right)  S^a \, \varepsilon_1
\, ,
\nonumber\\[3mm]
\delta S^a
&=&
\sqrt{2}\varepsilon_1\chi^a_L\,,
\nonumber\\[3mm]
\delta \zeta_L
&=&
i\sqrt{2} \left[ \left( \pt_0 -i\pt_3\right)z \right]\, \varepsilon_1\,,\qquad
z=x_{01}+ix_{02}\,,
\nonumber\\[3mm]
\delta z
&=&
\sqrt{2}\, \varepsilon_1\zeta_L\,,
\nonumber\\[3mm]
\delta \zeta_R
&=&
-2c\,\sqrt{2}\, \frac{\kappa}{m_W^2} \varepsilon_1\,\left[\chi^a_R\chi^a_L+
i\varepsilon^{abc}\,S^a\,\chi^b_R\chi^c_L\right]\,,
\label{02susy}
\eeqn
where 
$z$ is the complexified coordinate of the string center.
Here, much in the same way as in (\ref{susy2d}), we present for simplicity the SUSY
transformations generated only by the real parameter $\varepsilon_1$.
Note that we no longer have
$\varepsilon_2$-transformations in our worldsheet theory. These are 
broken by the \ntwoo deformation.

\section{Geometric formulation of the \boldmath\ntwoo hetero\-tic model}
\label{geom}

In this section we will derive the deformed $CP(1)$ model with no reference
to string worldsheet theory and  arguments in \cite{Edalati}, through an 
appropriately modified
superfield formalism. At the end we will present the Lagrangian of the heterotic $CP(N-1)$ model with arbitrary $N$.

\vspace{2mm}
\noindent
{\bf Warning:} In this section, unlike others, we use the Minkowski 
notation and normalize the kinetic term of $\zeta_R$ canonically.

\vspace{2mm}

To begin with, let us outline the standard geometric formulation of the
supersymmetric
$CP(N-1)$ sig\-ma model in 1+1 dimensions, $x^{\mu}=\{t,z\}$ \cite{BW, NSVZsigma}, with the subsequent deformation down to 
\ntwoo.

The target space is the $(N-1)$-dimensional K\"ahler manifold 
parametrized by the fields $\phi^{i},\,\phi^{\dagger\,\bar j}$, $\,i,\bar j=1,\ldots, N-1$,
which are the lowest components of the chiral and antichiral superfields 
\beqn
&&
\Phi^{i}(x^{\mu}+i\bar \theta \gamma^{\mu} \theta),\qquad \Phi^{\dagger\bar j}(x^{\mu}-i\bar \theta \gamma^{\mu} \theta)\,,
\nonumber\\[3mm]
&&
\Phi^{i} = \phi^i +\sqrt{2}\, \theta\,\psi^i + \theta^2\, F^i\,.
\eeqn
The Lagrangian is \cite{zum}
\begin{equation}
\label{eq:kinetic}
{\mathcal L} =\!\int\! {\rm d}^{4 }\theta K(\Phi, \Phi^{\dagger})
=G_{i\bar j} \big[\partial_\mu \phi^{\dagger\,\bar j}\, \partial_\mu\phi^{i}
+i\bar \psi^{\bar j} \gamma^{\mu} D_{\mu}\psi^{i}\big]
-\frac{1}{2}\,R_{i\bar jk\bar l}\,(\bar\psi^{\bar j}\psi^{i})(\bar\psi^{\bar l}\psi^{k})\,,
\end{equation}
where $K(\Phi, \Phi^{\dagger})$ is the K\"ahler potential,
$$
G_{i\bar j}=\frac{\partial^{2} K(\phi,\,\phi^{\dagger})}{\partial \phi^{i}\partial \phi^{\dagger\,\bar j}}
$$
 is the K\"ahler metric,
$R_{i\bar jk\bar l}$ is the Riemann tensor, 
$$ D_{\mu}\psi^{i}=
\partial_{\mu}\psi^{i}+\Gamma^{i}_{kl}\partial_{\mu} \phi^{k}\psi^{l}
$$
is the covariant derivative acting on the fermion field.   Our choice of the
gamma-matrices is summarized in Appendix B, along with some other definitions. 

A particular choice of the K\"ahler potential
\begin{equation}
\label{eq:kahler}
K =\frac{2}{g_{0}^{2}}\ln\left(1+\sum_{i,\bar j=1}^{ N-1}\Phi^{\dagger\,\bar j}\delta_{\bar j i}\Phi^{i}\right)
\end{equation}
is most common, it 
corresponds to the round Fubini--Study metric.
For $CP(N-1)$, the Ricci tensor $R_{i\bar j}$ is proportional to the metric,
\beq
\label{eq:RG}
R_{i\bar{j}} = \frac{g_{0}^2}{2}\,  N \, G_{i\bar{j}}\,,
\eeq
see also (\ref{640}).
 
The conserved  supercurrent is
\beq
J^{\mu}_{\alpha}=\sqrt{2}\,G_{i\bar j}\big[\partial_{\nu}\phi^{\dagger\bar j} \gamma^{\nu}\gamma^{\mu}\psi^{i}
\big]_{\alpha}\,.
\label{eq:superJ}
\eeq
In terms of the $R,L$ components it takes the form
\beqn
&&
J^0_R=J^1_R = \sqrt{2}\,G_{i\bar j}\big[\partial_{R}\phi^{\dagger\bar j} \big]\psi^{i}_R\,,
\nonumber\\[2mm]
&&
J^0_L= - J^1_L = \sqrt{2}\,G_{i\bar j}\big[\partial_{L}\phi^{\dagger\bar j} \big]\psi^{i}_L\,,
\eeqn
where 
\beq
\partial_L =\frac{\partial}{\partial t} +\frac{\partial}{\partial z}\,,\qquad
\partial_R =\frac{\partial}{\partial t} - \frac{\partial}{\partial z}\,,
\eeq
and
\beq
\psi =\left(\begin{array}{cc}
\psi_R \\
\psi_L
\end{array}
\right).
\eeq
Then the superconformal anomaly can be expressed as follows:
\beqn
&&
J_{\rm sc} \equiv \gamma_\mu J^\mu_{\,\alpha} =\frac{\sqrt 2}{2\pi} 
\,R_{i\bar j}\, \big[\partial_{\nu}\phi^{\dagger\bar j} \big]\, \gamma^{\nu}\psi^{i}\,,
\nonumber\\[3mm]
&&
J_{{\rm sc},R} =  \frac{- i \sqrt 2}{2\pi}\,R_{i\bar j}\,  \big[\partial_{R}\phi^{\dagger\bar j} \big]\psi^{i}_L\,,
 \qquad 
J_{{\rm sc},L} =  \frac{ i \sqrt 2}{2\pi}\,R_{i\bar j}\,  \big[\partial_{L}\phi^{\dagger\bar j} \big]\psi^{i}_R\,.
\label{anoma}
\eeqn

For what follows it is helpful to collect here explicit expressions in the case  of $CP(1)$. 
In this case a single complex field $\phi(t,z)$ serves as 
coordinate on the target space which is equivalent to $S^{2}$.
The K\"ahler potential $K$, the metric $G$, the Christoffel symbols $\Gamma,\,\bar\Gamma$ and the Ricci tensor $R$ are 
\beqn
&&
K\big|_{\theta=\bar\theta=0}=\frac{2}{g_{0}^{2}}\,\ln \chi\,, 
\nonumber
\\[2mm]
&&
G=G_{1\bar 1}=\partial_\phi\partial_{\phi^\dagger\,} 
K\big|_{\theta=\bar\theta=0}=
\frac{2}{g_{0}^2\,\chi^{2}}\,,
\nonumber
\\[2mm]
&&
\Gamma =\Gamma^{1}_{11} =- 2\, \frac{\phi^\dagger\,}{\chi}\,,\quad 
\bar\Gamma =\Gamma^{\bar 1}_{\bar 1\bar 1}=
- 2\, \frac{ \phi}{\chi}\,,
\nonumber
\\[2mm]
&&
 R \equiv  R_{1\bar 1}=-G^{-1}\!R_{1\bar 1 1\bar 1}=\frac{2}{\chi^2}\,,
\label{Atwo}
\eeqn
where we use the notation 
\beq
\chi \equiv 1+\phi\,\phi^\dagger\,.
\eeq
The Lagrangian of the conventional
$CP(1)$ model takes the form
\beqn
L_{CP(1)}
= G\,\Big\{
\partial_\mu \phi^{\dagger}\, \partial^\mu\phi
+i\bar \psi \gamma^{\mu}\partial_{\mu}\psi
-\frac{2i}{\chi}\, \phi^{\dagger}\partial_{\mu}\phi\,\bar\psi \gamma^{\mu}\psi+
\frac{1}{\chi^{2}}\,(\bar \psi \psi)^{2}\Big\}\,.
\label{cpone}
\eeqn
 In terms of the components $\psi_{L,R}$,
the Lagrangian (\ref{cpone}) can be rewritten as
\beqn
&&
L_{\,  CP(1)}= G\, \left\{\rule{0mm}{5mm}
\partial_\mu \phi^\dagger\, \partial^\mu\phi  
+\frac{i}{2}\big(\psi_L^\dagger\!\stackrel{\leftrightarrow}{\partial_R}\!\psi_L 
+ \psi_R^\dagger\!\stackrel{\leftrightarrow}{\partial_L}\!\psi_R
\big)\right.
\nonumber
\\[3mm] 
&&
-\frac{i}{\chi}\,  \big[\psi_L^\dagger \psi_L
\big(\phi^\dagger \!\stackrel{\leftrightarrow}{\partial_R}\!\phi
\big)+ \psi_R^\dagger\, \psi_R
\big(\phi^\dagger\!\stackrel{\leftrightarrow}{\partial_L}\!\phi
\big)
\big]
-
\frac{2}{\chi^2}\,\psi_L^\dagger\,\psi_L \,\psi_R^\dagger\,\psi_R
\Big\}\,,
\label{Aone}
\eeqn
where 
$\partial_{L,R} = \partial /\partial t \pm \partial /\partial z$, see Appendix B.
It is not difficult to check that it has \ntwot supersymmetry.

Now, let us introduce a heterotic deformation, due to the
right-handed field $\zeta_R$, which transforms the above 
\ntwot model into \ntwoo$\!,$
\beqn
&&
L_{{\rm heterotic}}= 
\zeta_R^\dagger \, i\partial_L \, \zeta_R  + 
\left[\gamma \, \zeta_R  \,R\,  \big( i\,\partial_{L}\phi^{\dagger} \big)\psi_R
+{\rm H.c.}\right] -g_0^2|\gamma |^2 \left(\zeta_R^\dagger\, \zeta_R
\right)\left(R\,  \psi_L^\dagger\psi_L\right)
\nonumber
\\[4mm]
&&
+
G\, \left\{\rule{0mm}{5mm}
\partial_\mu \phi^\dagger\, \partial^\mu\phi  
+\frac{i}{2}\big(\psi_L^\dagger\!\stackrel{\leftrightarrow}{\partial_R}\!\psi_L 
+ \psi_R^\dagger\!\stackrel{\leftrightarrow}{\partial_L}\!\psi_R
\big)\right.
\nonumber
\\[4mm] 
&&
-\frac{i}{\chi}\,  \big[\psi_L^\dagger \psi_L
\big(\phi^\dagger \!\stackrel{\leftrightarrow}{\partial_R}\!\phi
\big)+ \psi_R^\dagger\, \psi_R
\big(\phi^\dagger\!\stackrel{\leftrightarrow}{\partial_L}\!\phi
\big)
\big]
-
\frac{2(1- g_0^2 |\gamma |^2)}{\chi^2}\,\psi_L^\dagger\,\psi_L \,\psi_R^\dagger\,\psi_R
\Big\}\,,
\label{AAone}
\eeqn
where $R$ stands for the Ricci tensor, see Eq.~(\ref{Atwo}).
One can obtain the deformed Lagrangian (\ref{AAone}) as follows.
Introduce the operators
\beqn
{\mathcal B} &=& \left\{\,\zeta_R (x^\mu +i\bar\theta\gamma^\mu\theta) +\sqrt{2}\theta_R{\mathcal F}
\right\} \theta_L^\dagger\,,
\nonumber\\[2mm]
{\mathcal B}^\dagger &=&\theta_L \left\{\,\zeta_R^\dagger (x^\mu - i\bar\theta\gamma^\mu\theta) +\sqrt{2}\theta_R^\dagger {\mathcal F}^\dagger
\right\} \,.
\label{AAA}
\eeqn
Since $\theta_L$ and $\theta_L^\dagger$ enter in Eq.~(\ref{AAA})  explicitly,
${\mathcal B}$ and ${\mathcal B}^\dagger$ are {\em not} superfields with regards
to the supertransformations with parameters $\epsilon_L,\, \epsilon_L^\dagger$.
These supertransformations are absent in the heterotic model.
Only those survive which are associated with  $\epsilon_R,\, \epsilon_R^\dagger$. 
Note that ${\mathcal B}$ and ${\mathcal B}^\dagger $
are superfields with regards to the latter. Here $\epsilon_R$ is related to the real parameter
of supersymmetry transformations we dealt with in previous sections as ${\rm Re}\,\epsilon_R= \varepsilon_1$.

It is convenient to introduce a shorthand for the chiral coordinate
\beq
\tilde{x}^\mu =x^\mu + i\bar\theta\gamma^\mu\theta\,.
\eeq
Then the transformation laws with the parameters $\epsilon_R,\, \epsilon_R^\dagger$
are as follows:
\beq
\delta\theta_R =\epsilon_R\,,\quad \delta\theta_R^\dagger =\epsilon_R^\dagger
\,,\quad \delta\tilde{x}^0 = 2i \epsilon_R^\dagger\theta_R
\,,\quad \delta\tilde{x}^1 = 2i \epsilon_R^\dagger\theta_R\,.
\eeq
With respect to such supertransformations, ${\mathcal B}$ and ${\mathcal B}^\dagger$ are  superfields.
Indeed,
\beq
\delta\zeta_R = \sqrt{2}\, {\mathcal F} \,\epsilon_R\,,\quad  \delta {\mathcal F} =\sqrt{2}\,i\left(\partial_L
\zeta_R\right) \epsilon_R^\dagger\,,
\eeq
plus Hermitian conjugate transformations. To convert
$L_{CP(1)}$ into $L_{{\rm heterotic}}$
we add to $L_{CP(1)}$ the following terms:
\beqn
\Delta L =  \int d^4\theta \left\{-2\, {\mathcal B}^\dagger \,{\mathcal B}
+ \left[{g_0^2}\, {\sqrt 2}\, \gamma\,  {\mathcal B}\, K +{\rm H.c.}\right]
\right\},
\label{exte}
\eeqn
where $\gamma$ is generally speaking a complex constant.
For simplicity we will assume $\gamma$ to be real.
Thus, we obviously deal here with a single deformation parameter.
Its relation to $\alpha$ is discussed in Sect.~\ref{02} (see Eq.~(\ref{gamal}))
while the relation to the bulk theory parameters in Sect.~\ref{crossterm}.
To derive  Eq.~(\ref{gamal}) it is sufficient to compare
the $\zeta_R\partial_L\phi^\dagger\psi_R$ term in Eq.~(\ref{AAone})
with the $\chi_R^a\partial_L S^a (\zeta_R +\bar\zeta_R)$ term in
Eq.~(\ref{02o3p}). 

First, let us check that the extra term (\ref{exte})
preserves invariance on the target space.
Indeed, the invariance under the U(1) transformation of the
superfields $\Phi,\,\Phi^\dagger$,
\beq
\Phi\to i\delta\,\Phi\,,\quad \Phi^\dagger\to- i\delta\,\Phi^\dagger\,.
\eeq
is  obvious. Two other rotations on the sphere manifest themselves in nonlinear transformations with a complex parameter $\beta$,
\beq
\Phi \to \beta +\beta^* \Phi^2\,,\quad \Phi^\dagger \to \beta^* +\beta 
\left( \Phi^\dagger\right)^2\,.
\eeq
Under these transformations
\beq
\delta K = \frac{2}{g_0^2}\, \left(\beta^*\,\Phi + \beta\,\Phi^\dagger\right).
\eeq
It is not difficult to see that
\beq
\int d^4\theta\, {\mathcal B}\, \delta K = 0\,.
\eeq
In other words, even before performing the component decomposition
we are certain that the term (\ref{exte})
is invariant on the target space of the $CP(1)$ model. Needless to say, it is \ntwoo
invariant by construction.

As usual, the ${\mathcal F}$ term enters without derivatives and can be eliminated
by virtue of equations of motion,
\beq
{\mathcal F} = -2\, \gamma^*\, \chi^{-2}\, \psi_R^\dagger\,\psi_L\,,\quad 
{\mathcal F}^\dagger = -2\, \gamma\, \chi^{-2}\, \psi_L^\dagger\,\psi_R\,.
\label{ft}
\eeq    
This is responsible for the change of the standard coefficient in front
of the four-fermion term (the last line in Eq.~(\ref{AAone})). 
As for the target space structure of this coefficient, it is proportional to
the curvature tensor of $CP(1)$.

In addition, the $F$ terms of the superfields $\Phi,\,\,\Phi^\dagger$ also change.
If before the deformation e.g. $F=(i/2)\,\Gamma \,\psi\,\gamma^0\,\psi$, after
the deformation
\beq
F =\frac{i}{2}\,\Gamma \,\psi\,\gamma^0\,\psi - g_0^2 \,\gamma\,  
\psi_L \, \zeta_R^\dagger\,,
\label{ftp}
\eeq
plus the Hermitian conjugated expression for $F^\dagger$. As a result, we get the last term in the first line in Eq.~(\ref{AAone}).

If we take into account the relation between $\gamma$ and the bulk theory
parameters we can conclude that $g_0^2|\gamma |^2 <1$.

The last issue to be discussed in this section is the change of the supercurrent. The supercurrent in the conventional $CP(1)$ model is given in Eq.~(\ref{anoma}). When we deform the
model the supercurrent acquires  extra terms associated with the 
$\mathcal F$, $F$ terms in Eqs. (\ref{ft}) and (\ref{ftp}). This term is proportional to
$\gamma \left\{R \,  \psi_R^\dagger\,\psi_L\right\} \zeta_R^\dagger$, 
and its Hermitian conjugate, of course. Assume that $\gamma\ll 1$. 
Then the expression in the braces can be evaluated in the undeformed
$CP(1)$ model. As well known (see e.g. \cite{NSVZsigma}),
a nonvanishing bifermion condensate $\langle R \,  \psi_R^\dagger\,\psi_L\rangle 
\sim \pm \Lambda$ develops in this model ($\Lambda$ is the scale parameter)
labeling two distinct vacua. Thus, the additional terms in the supercurrent
emerging in the deformed theory (at small $\gamma$)
have the form
\beq
\Delta J_{\rm sc} =\gamma \, \langle R \,  \psi_R^\dagger\,\psi_L\rangle \,
\zeta_R^\dagger\,,\quad \Delta J_{\rm sc}^\dagger = \gamma \, \langle R \,  \psi_L^\dagger\,\psi_R\rangle \,
\zeta_R\,.
\label{golz}
\eeq
Since $\zeta_R$ is strictly massless Eq.~(\ref{golz})
clearly demonstrates that $\zeta_R$ is a Goldstino, with the residue $\langle R \,  \psi_R^\dagger\,\psi_L\rangle$. Supersymmetry is spontaneously broken, with the
vacuum energy 
\beq
{\mathcal E}_{\rm vac} = |\gamma|^2\,
\left|\langle R \,  \psi_R^\dagger\,\psi_L\rangle
\right|^2
\eeq
times a numerical factor, one and the same for both vacua. In the accompanying paper
\cite{ACC} we will obtain a nonvanishing ${\mathcal E}_{\rm vac}$ for arbitrary values of $\gamma$ in heterotically deformed $CP(N-1)$ models
using large $N$ expansion.  The very possibility of the spontaneous supersymmetry breaking
is due to the fact that Witten's index $I_W$ of the
deformed theory vanishes, in sharp contradistinction with the 
undeformed conventional
\ntwot model where $I_W =N$ \cite{WI}. Details of this statement are discussed in 
Appendix C.

Given the geometric representation of the deformed $CP(1)$ model
(\ref{AAone}) one can suggest a generalization for arbitrary $N$
(i.e. the \ntwoo deformed $CP(N-1)$ model), 
\beqn
L_{{\rm heterotic}} && 
= 
\zeta_R^\dagger \, i\partial_L \, \zeta_R  + 
\left[\gamma\, g_0^2 \, \zeta_R  \, G_{i\bar j}\,  \big( i\,\partial_{L}\phi^{\dagger\,\bar j} \big)\psi_R^i
+{\rm H.c.}\right]
\nonumber
\\[4mm]
&&
 -g_0^4\, |\gamma |^2 \,\left(\zeta_R^\dagger\, \zeta_R
\right)\left(G_{i\bar j}\,  \psi_L^{\dagger\,\bar j}\psi_L^i\right)
\nonumber
\\[4mm]
&&
+G_{i\bar j} \big[\partial_\mu \phi^{\dagger\,\bar j}\, \partial_\mu\phi^{i}
+i\bar \psi^{\bar j} \gamma^{\mu} D_{\mu}\psi^{i}\big]
\nonumber
\\[4mm]
&&
- \frac{g_0^2}{2}\left( G_{i\bar j}\psi^{\dagger\, \bar j}_R\, \psi^{ i}_R\right)
\left( G_{k\bar m}\psi^{\dagger\, \bar m}_L\, \psi^{ k}_L\right)
\nonumber
\\[4mm]
&&
+\frac{g_0^2}{2}\left(1-2g^2_0|\gamma|^2\right)
\left( G_{i\bar j}\psi^{\dagger\, \bar j}_R\, \psi^{ i}_L\right)
\left( G_{k\bar m}\psi^{\dagger\, \bar m}_L\, \psi^{ k}_R\right)\,,
\label{cpn-1g}
\eeqn
where we used the fact that
for the above K\"ahler metric
\beq
R_{i\bar{j} k\bar{m}} = - \frac{g_0^2}{2}\left(G_{i\bar{j}}G_{k\bar{m}} +
G_{i\bar{m}}G_{k\bar{j}}
\right)\,.
\label{640}
\eeq
We assume that $g^{-2}_0$ is proportional to $N$
while $|\gamma |^2 g_0^2$ has no $N$ dependence.
Note that the term in the fourth line is absent in \cite{Edalati}.

\section{From the bulk  \boldmath\none theory to the heterotic
deformation of the $CP(1)$ model on the  worldsheet}
\label{N1zeromodes}
\setcounter{equation}{0}

In this section we will obtain fermion zero modes on the non-Abelian string
in the bulk theory with the deformation term (\ref{msuperpotbr}). 
This will allow us, eventually, 
to calculate the bifermion cross-term $\bar\chi \partial_L S \zeta_R$
directly from the bulk theory and thus  determine the parameter
$\kappa$ in terms of the bulk theory parameters. This will completely fix the heterotic
model since the overall structure of the Lagrangian is dictated by \ntwoo supersymmetry. 

\subsection{Fermion zero modes in \boldmath\none theory}

We start from supertranslational fermion zero modes on the string. Superorientational
zero modes
for the bulk deformation (\ref{msuperpotbr}) were determined in \cite{SYnone}
and we just quote the results. With \ntwo supersymmetry broken, we have only
two complex SUSY transformations left. They are generated by parameters $\epsilon^{11}$
and $\epsilon^{21}$, see Sect.~\ref{fermzeromodes}. The latter transformation acts
on the string solution trivially while the former gives two unmodified fermion 
supertranslational zero modes in (\ref{tzmodes}) proportional to 
$\zeta_L$. At the same time,
the fields $\lambda^{22}$ and $\bar{\tilde{\psi}}_{\dot{1}}$ proportional
to $\zeta_R$ are modified. To find them we explicitly solve below the Dirac 
equations (\ref{dirac1}) and (\ref{dirac2}).

It is easy to check that the modified fermion fields $\lambda^{22}$ and 
$\bar{\tilde{\psi}}_{\dot{1}}$ can be written in  
the following form:
\beqn
\lambda^{22}
&=&
\lambda_{s0}\,\zeta_R +\lambda_{s1}\,\frac{x_1+ix_2}{r}\,\bar{\zeta}_R\, ,
\nonumber\\[4mm]
\lambda^{a22}
&=&
\lambda_{t0}\,S^a\,\zeta_R +\lambda_{t1}\,S^a\frac{x_1+ix_2}{r}\,
\bar{\zeta}_R\, ,
\nonumber\\[4mm]
\bar{\tilde{\psi}}_{\dot{1}}
&=&
\frac12\,\frac{x_1-ix_2}{r}\,\left(\psi_{s0}+\tau^a S^a\psi_{t0}\right)\,
\zeta_R +
\frac12\,\left(\psi_{s1}+\tau^a S^a\psi_{t1}\right)\,\bar{\zeta}_R\, ,
\nonumber\\[1mm]
\label{ftprofile}
\eeqn
where we introduced four profile functions $\lambda(r)$, and four functions $\psi(r)$
parameterizing the fermion fields $\lambda^{22}$ and $\bar{\tilde{\psi}}_{\dot{1}}\,,$
respectively. The subscripts $s$ and $t$ label the singlet and triplet profile
functions with respect to the unbroken global SU(2)$_{C+F}$.

Substituting (\ref{ftprofile}) in the Dirac equations (\ref{dirac1}) and 
(\ref{dirac2}) we get  equations for the fermion profile functions
for $\lambda$ fermions
\beqn
&-&\frac{d}{dr}\lambda_{s0}
+i\frac{g^2_1}{2\sqrt{2}}\,\left[(\phi_1+\phi_2)\,\psi_{s0}+(\phi_1-\phi_2)\,\psi_{t0}
\right]
 -g^2_1\mu\,\lambda_{s1}=0\,,
\nonumber\\[3mm]
&-&\frac{d}{dr}\lambda_{s1}-\frac1r\lambda_{s1}
+i\frac{g^2_1}{2\sqrt{2}}\,\left[(\phi_1+\phi_2)\,\psi_{s1}+(\phi_1-\phi_2)\,\psi_{t1}
\right]
 -g^2_1\mu\,\lambda_{s0}=0\,,
\nonumber\\[3mm]
&-&\frac{d}{dr}\lambda_{t0}
+i\frac{g^2_2}{2\sqrt{2}}\,\left[(\phi_1-\phi_2)\,\psi_{s0}+(\phi_1+\phi_2)\,\psi_{t0}
\right]
 -g^2_2\mu\,\lambda_{t1}=0\,,
\nonumber\\[3mm]
&-&\frac{d}{dr}\lambda_{t1}-\frac1r\lambda_{t1}
+i\frac{g^2_2}{2\sqrt{2}}\,\left[(\phi_1-\phi_2)\,\psi_{s1}+(\phi_1+\phi_2)\,\psi_{t1}
\right]
 -g^2_2\mu\,\lambda_{t0}=0\,,
\nonumber\\[1mm]
\label{lambdaeqs}
\eeqn
and for $\psi$ fermions
\beqn
&&\frac{d}{dr}\psi_{s0} +\frac1r\psi_{s0}-\frac{f}{2r}\psi_{s0}-\frac{f_3}{2r}\psi_{t0}
+\frac{i}{\sqrt{2}}\,\left[(\phi_1+\phi_2)\,\lambda_{s0}+(\phi_1-\phi_2)\,\lambda_{t0}
\right] =0\,,
\nonumber\\[3mm]
&&\frac{d}{dr}\psi_{t0} +\frac1r\psi_{t0}-\frac{f}{2r}\psi_{t0}-\frac{f_3}{2r}\psi_{s0}
+\frac{i}{\sqrt{2}}\,\left[(\phi_1-\phi_2)\,\lambda_{s0}+(\phi_1+\phi_2)\,\lambda_{t0}
\right] =0\,,
\nonumber\\[3mm]
&&\frac{d}{dr}\psi_{s1} -\frac{f}{2r}\psi_{s1}-\frac{f_3}{2r}\psi_{t1}
+\frac{i}{\sqrt{2}}\,\left[(\phi_1+\phi_2)\,\lambda_{s1}+(\phi_1-\phi_2)\,\lambda_{t1}
\right] =0\,,
\nonumber\\[3mm]
&&\frac{d}{dr}\psi_{t1} -\frac{f}{2r}\psi_{t1}-\frac{f_3}{2r}\psi_{s1}
+\frac{i}{\sqrt{2}}\,\left[(\phi_1-\phi_2)\,\lambda_{s1}+(\phi_1+\phi_2)\,\lambda_{t1}
\right] =0\,.
\nonumber\\[1mm]
\label{psieqs}
\eeqn
We will solve these equations below in two limits, for small and 
large values of the deformation parameter $\mu$.

As was already mentioned, the superorientational zero modes in the theory with 
deformation (\ref{msuperpotbr}) were found in \cite{SYnone}. They have the form
\beqn
\lambda^{a22}
&=&
\frac{x_1+ix_2}{r}\,\lambda_{+}(r)\,\left[\chi_R^a-
i\varepsilon^{abc}S^b\chi_R^c\right] + \lambda_{-}(r)\,\left[\chi_R^a+
i\varepsilon^{abc}S^b\chi_R^c\right]\, ,
\nonumber\\[4mm]
\bar{\tilde{\psi}}^{kA}_{\dot{1}}
&=&
\psi_{+}(r)\,
\left(\frac{\tau^a}{2}\right)^{kA}\,
\left[\chi_R^a-
i\varepsilon^{abc}S^b\chi_R^c\right]
\nonumber\\[4mm]
&+&
\frac{x_1-ix_2}{r}\,\psi_{-}(r)\,
\left(\frac{\tau^a}{2}\right)^{kA}\,
\left[\chi_R^a+
i\varepsilon^{abc}S^b\chi_R^c\right],
\label{fprofile}
\eeqn
where we introduced four profile functions $\lambda_{\pm}$ and $\psi_{\pm}$
parameterizing the fermion fields $\lambda^{22}$ and $\bar{\tilde{\psi}}_{\dot{1}}$.
The functions $\lambda_{+}$ and $\psi_{+}$ are expandable
in even powers of
$\mu$ while the functions  $\lambda_{-}$ and $\psi_{-}$    in
odd powers of $\mu$.

Substituting (\ref{fprofile}) into the Dirac equations (\ref{dirac1}),
(\ref{dirac2}) we get following equations for the fermion profile functions:
\beqn
&&\frac{d}{dr}\psi_{+} -\frac1{2r}(f-f_3)\psi_{+}+i\sqrt{2}\,\phi_1\,
\lambda_{+}=0\, ,
\nonumber\\[3mm]
&-&\frac{d}{dr}\lambda_{+}-\frac1r\lambda_{+}+\frac{f_3}{r}\lambda_{+}
+i\frac{g^2_2}{\sqrt{2}}\,\phi_1\,\psi_{+} +g^2_2\mu\,\lambda_{-}=0\,,
\nonumber\\[3mm]
&&\frac{d}{dr}\psi_{-} +\frac1r\psi_{-}
-\frac1{2r}(f+f_3)\psi_{-}+i\sqrt{2}\,\phi_2\,\lambda_{-}=0,
\nonumber\\[3mm]
&-&\frac{d}{dr}\lambda_{-}-\frac{f_3}{r}\lambda_{-}
+i\frac{g^2_2}{\sqrt{2}}\,\phi_2\,\psi_{-} +g^2_2\mu\,\lambda_{+}=0\,.
\label{fermeqs}
\eeqn

\subsection{Small-\boldmath{$\mu$} limit}

In terms of the profile functions introduced in (\ref{ftprofile}) the 
undeformed translational fermion zero modes (\ref{tzmodes}) of the \ntwot model
model are
\beqn
\lambda_{s0}
&=&
ig^2_1\,(\phi_1^2+\phi_2^2-2\xi)\,
+O(\mu^2)\,,\quad
\lambda_{t0}=ig^2_2\,(\phi_1^2-\phi_2^2)\,
+O(\mu^2),
\nonumber\\[3mm]
\psi_{s0}
&=&
\frac{\sqrt{2}}{r}\left[(f+f_3)\phi_1+(f-f_3)\phi_2\right]+O(\mu^2)\,,
\nonumber\\[3mm]
\psi_{t0}
&=&
\frac{\sqrt{2}}{r}\left[(f+f_3)\phi_1-(f-f_3)\phi_2\right]+O(\mu^2)\,.
\label{tzeroorder}
\eeqn
The profile functions $\lambda_0$ and $\psi_0$ can be expressed as series in even
powers of $\mu$. 

On the other hand, the functions $\lambda_1$ and $\psi_1$ are expandable in
odd powers of $\mu$. The Dirac equations (\ref{lambdaeqs}) and (\ref{psieqs}) can be 
easily solved for these functions, to the leading order in $\mu$. We consider
$\mu$-dependent terms in the second
and the  last equations in (\ref{lambdaeqs}) as perturbations 
and substitute there the solutions (\ref{tzeroorder}). Then we have
\beqn
&&
\lambda_{s1}=-\frac{g^2\mu}{2}r\,\lambda_{s0}
+O(\mu^3),\qquad
\lambda_{t1}=-\frac{g^2\mu}{2}r\,\lambda_{t0}\,
+O(\mu^3),
\nonumber\\[3mm]
&&
\psi_{s1}=-\frac{g^2\mu}{2}r\,\psi_{s0}+O(\mu^3)\, , \qquad
\psi_{t1}=-\frac{g^2\mu}{2}r\,\psi_{t0}+O(\mu^3)\, ,
\label{tfirstorder}
\eeqn
where we put $g_1=g_2\equiv g$ for simplicity.

The behavior of the superorientational fermion zero modes in the small-$\mu$ limit
was obtained in \cite{SYnone}. The  leading contributions to the $\mu$-even profile functions are
\beq
\lambda_{+}=\frac{i}{\sqrt{2}}\frac{f_3}{r}\frac{\phi_1}{\phi_{2}}
+O(\mu^2),\,\,\,\,\,
\psi_{+}=\frac{1}{2\phi_2}\left(\phi_1^2-\phi_2^2\right)+O(\mu^2)\,,
\label{zeroorder}
\eeq
see Eq.~(\ref{ozmodes}).

Substituting Eq.~(\ref{zeroorder}) into the last equation in (\ref{fermeqs})
we can solve for the leading contributions to the $\mu$-odd profile functions.
In this way we get \cite{SYnone}
\beqn
\lambda_{-}
&=&
g_2^2\mu\,\frac{i}{2\sqrt{2}}\,\left[(f_3-1)\frac{\phi_2}{\phi_1}
+\frac{\phi_1}{\phi_2}\right]+O(\mu^3)\,,
\nonumber\\[3mm]
\psi_{-}
&=&
g_2^2\mu\frac{r}{4\phi_1}\left(\phi_1^2-\phi_2^2\right)+
O(\mu^3)\, .
\label{firstorder}
\eeqn
Using the boundary conditions (\ref{fbc}) and (\ref{phibc})
for the string profile functions  it is easy to check that these solutions
vanish at $r\to\infty$ and are nonsingular at $r=0$.

\subsection{Large-\boldmath{$\mu$} limit}

Now let us dwell on the limit of large $\mu$. As was explained in Sect.~\ref{bulk}, the
fields  $a$, $a^a$ (as well as their fermion counterparts $\lambda^{\alpha 2}$,
$\lambda^{a\alpha 2}$)
become heavy and can be integrated out. The low-energy theory 
 contains  massive U(1) and SU(2) gauge multiplets, with mass
(\ref{phmass}) and (\ref{wmass}),
 and four  chiral light multiplets, with mass
\beq
m_L\equiv m^{-}_{{\rm SU}(2)} =m^{-}_{{\rm U}(1)}=\frac{\xi}{\mu}\, ,
\label{lmass}
\eeq
see Eq.~(\ref{light}).

Integrating out heavy fields can be performed directly in the superpotential,
as in \cite{KSS,GVY,VY}, or in the component Lagrangian. To this end one drops
the kinetic terms for all heavy fields
and solves algebraic equations for these fields. We do it in the
fermion sector of the theory in the Dirac equations  (\ref{dirac1})
for $\lambda^{\alpha 2}$ and $\lambda^{a\alpha 2}$. First we consider  
supertranslational zero modes. The large-$\mu$ limit for superorientational modes was
considered in \cite{SYnone} and we just quote the corresponding results.

More exactly, we  get  expressions for
the $\lambda$-profile functions in terms of the $\psi$-profile functions
from  equations in (\ref{psieqs}). Namely,
\beqn
\lambda_{s0}+\lambda_{t0}
&=&
\frac{i\sqrt{2}}{\phi_1}\left[\frac{d}{dr}\psi_{+0} + \frac1r\psi_{+0}
-\frac1{2r}(f+f_3)\psi_{+0}\right],
\nonumber\\[3mm]
\lambda_{s0}-\lambda_{t0}
&=&
\frac{i\sqrt{2}}{\phi_2}\left[\frac{d}{dr}\psi_{-0} + \frac1r\psi_{-0}
-\frac1{2r}(f-f_3)\psi_{-0}\right],
\nonumber\\[3mm]
\lambda_{s1}+\lambda_{t1}
&=&
\frac{i\sqrt{2}}{\phi_1}\left[\frac{d}{dr}\psi_{+1}  
-\frac1{2r}(f+f_3)\psi_{+1}\right],
\nonumber\\[3mm]
\lambda_{s1}-\lambda_{t1}
&=&
\frac{i\sqrt{2}}{\phi_2}\left[\frac{d}{dr}\psi_{-1}  
-\frac1{2r}(f-f_3)\psi_{-1}\right],
\label{tlambdapsi}
\eeqn
where
\beq
\psi_{\pm 0}=\frac12(\psi_{s0} \pm\psi_{t0}), \qquad 
\psi_{\pm 1}=\frac12(\psi_{s1} \pm\psi_{t1}).
\label{psipm}
\eeq
Dropping the kinetic terms for
$\lambda$'s  in Eqs.~(\ref{lambdaeqs}) and
substituting (\ref{tlambdapsi}) in these equations we arrive at
\beqn
&&\frac{d}{dr}\psi_{+0} +\frac1r\,\psi_{+0}
-\frac1{2r}(f+f_3)\psi_{+0}-m_L\,\frac{\phi_1^2}{\xi}
\,\psi_{+1}=0\, ,
\nonumber\\[3mm]
&&\frac{d}{dr}\psi_{-0} +\frac1r\,\psi_{-0}
-\frac1{2r}(f-f_3)\psi_{-0}-m_L\,\frac{\phi_2^2}{\xi}
\,\psi_{-1}=0\, ,
\nonumber\\[3mm]
&&\frac{d}{dr}\psi_{+1} 
-\frac1{2r}(f+f_3)\psi_{+1}-m_L\,\frac{\phi_1^2}{\xi}
\,\psi_{+0}=0\, ,
\nonumber\\[3mm]
&&\frac{d}{dr}\psi_{-1} 
-\frac1{2r}(f-f_3)\psi_{-1}-m_L\,\frac{\phi_2^2}{\xi}
\,\psi_{-0}=0\, ,
\label{largemupsieqs}
\eeqn
where $m_L$ is the light mass given in Eq.~(\ref{lmass}).

Now please observe that the long-range tails of the solutions to these equations are
determined by the small mass $m_L$, while the string profile functions
$f$ and $f_3$ are important at much smaller distances $r\sim 1/g\sqrt{\xi}$
(we assume that  both coupling constants are of the same order, $g_1\sim g_2\sim g$).
This key observation allows us to solve Eqs.~(\ref{largemupsieqs})
analytically. We will treat separately two domains,
\beqn
&&
 \mbox{Large}\,\,\, r\,,\quad  r\gg 1/g\sqrt{\xi},
\label{larger}
\\[3mm]
&&
\mbox{Intermediate} \,\,\, r\,,\quad  r\lsim 1/g\sqrt{\xi}\,.
\label{interr}
\eeqn

In the large-$r$  domain (\ref{larger})
we can drop the terms in (\ref{largemupsieqs}) containing $f$ and $f_3$ and use the
last two  equations in (\ref{largemupsieqs})
to express $\psi_{0}$ in terms of $\psi_{1}$. We then get
\beq
\psi_{+0}= \frac1{m_L}\frac{d}{dr}\psi_{+1}\, , \qquad
\psi_{-0}= \frac1{m_L}\frac{d}{dr}\psi_{-1}\, .
\label{psi01}
\eeq
Substituting this into the first two  equations in (\ref{largemupsieqs}) we obtain
\beqn
&&
\frac{d^2}{dr^2}\psi_{+1}+\frac1r\frac{d}{dr}\psi_{+1}-m_L^2\psi_{+1}=
0\, ,
\nonumber \\[3mm]
&&
\frac{d^2}{dr^2}\psi_{-1}+\frac1r\frac{d}{dr}\psi_{-1}-m_L^2\psi_{-1}=
0\, .
\label{psi1eq}
\eeqn
These are well-known equations for free fields with mass $m_L$
in the radial coordinates. Their solutions 
are\,\footnote{Equation~(\ref{psi1eq}) determines
the profile function $\psi_{1}$ up to an overall normalization constant.
This constant is included in the normalization of
the two-dimensional fermion
field $\zeta_R$. We will discuss this normalization in the next section.}
\beqn
&&
\psi_{+1}=m_L \sqrt{\xi}\, K_0(m_L r),
\nonumber \\[3mm]
&&
\psi_{-1}=m_L \sqrt{\xi}\, K_0(m_L r),
\label{psi1}
\eeqn
where $K_0 (x)$ is the imaginary argument Bessel function (the McDonald function). At infinity it falls off exponentially,
\beq
K_0(x)\sim \frac{e^{-x}}{\sqrt{x}}\,,
\eeq
while at $x\to 0$ it has the
logarithmic behavior,
\beq
K_0(x)\sim \ln{\frac1x}\, .
\label{log}
\eeq
Taking into account (\ref{psi01}) we get the solutions for
the fermion profile functions at $r\gg 1/m_0$,
\beqn
&&
\psi_{+1}\sim m_L\sqrt{\xi}\; K_0(m_L r)\,,\qquad 
\psi_{+0}\sim \sqrt{\xi}\; \frac{d}{dr}K_0(m_L r)\,,
\nonumber \\[3mm]
&&
\psi_{-1}\sim m_L\sqrt{\xi}\; K_0(m_L r)\,,\qquad 
\psi_{-0}\sim \sqrt{\xi}\; \frac{d}{dr}K_0(m_L r)\, .
\label{largerpsi}
\eeqn
In particular, at $1/(g\sqrt{\xi})\ll r\ll 1/m_L$ we have
\beq
\psi_{\pm 1}\sim m_L\sqrt{\xi}\, \ln\, {\frac1{m_L r}}\,,
\qquad
\psi_{\pm 0}\sim\sqrt{\xi}\;\frac1r  \,.
\label{psizero}
\eeq
In Sect.~\ref{crossterm} we will see that the long-range $1/r$ tails of the profile
functions  give the leading (logarithmic) contributions to
 normalization integrals in front of kinetic terms for two-dimensional
fermion fields on the string worldsheet (\ref{02o3}).
Therefore, it is essential to determine the precise combination of the profile functions
$\psi_{s0}$ and $\psi_{t0}$ which has this long-range behavior. To this end we match
the behavior at large $r$ given by (\ref{largerpsi}) and  (\ref{psizero}) with the 
behavior of these functions at intermediate $r$, in the domain (\ref{interr}).

In this domain we neglect
small mass terms in (\ref{largemupsieqs}). We then arrive at
\beqn
&&\frac{d}{dr}\psi_{+0} +\frac1r\,\psi_{+0}
-\frac1{2r}(f+f_3)\psi_{+0}=0\, ,
\nonumber\\[3mm]
&&\frac{d}{dr}\psi_{-0} +\frac1r\,\psi_{-0}
-\frac1{2r}(f-f_3)\psi_{-0}=0\, ,
\label{interrpsieqs}
\eeqn
were we restrict ourselves to the equations for the profile functions $\psi_0$ which
have the long-range $1/r$ tails.
These equations are identical to those for the string profile
functions, see Eq.~(\ref{foe}). Therefore, their solutions are known,
\beq
\psi_{+0}=c_1\frac{\phi_1}{r}\,,
\qquad
\psi_{-0}=c_2 \frac{\phi_2}{r}\,.
\label{spsi}
\eeq
Since the profile function $\phi_2\sim {\rm const}$ at $r\to 0$ the coefficient $c_2$
above should vanish, $$c_2=0\,,$$ otherwise the function $\psi_{-0}$ would be 
singular at $r=0$. The profile function $\phi_1\sim r$ at $r\to 0$.
Therefore, finally we get
\beq
\psi_{s0}=\psi_{t0}=\frac{\phi_1}{r}
\label{intrpsi0}
\eeq
at intermediate $r$, and 
\beq
\psi_{s0}=\psi_{t0}=-\sqrt{\xi} \;\frac{d}{dr}K_0(m_L r)
\label{largerpsi0}
\eeq
at large $r$,
where we include the constant  $c_1$ in the normalization of the two-dimensional
field $\zeta_R$, see Sect.~\ref{crossterm}.

A similar procedure was used in \cite{SYnone} to determine the long-range
tails of the superorientational zero modes. It turns out that the profile function
$\psi_{-}$ (see (\ref{fprofile}))
has the $1/r$ long-range tail. The result obtained in \cite{SYnone} for this function
is
\beq
\psi_{-}=\frac{\phi_1}{r\sqrt{\xi}}
\label{intrpsi-}
\eeq
at intermediate $r$, and 
\beq
\psi_{-}= - \frac{d}{dr}K_0(m_L r) 
\label{largerpsi-}
\eeq
at large $r$.

Note, that the main
feature of the fermion zero modes described above is the presence of the 
long-range tails determined
by the {\em small} mass $m_L$. Neither the bosonic string solution
(\ref{sna}) nor other supertranslational and superorientational fermion zero modes
determined by \none supersymmetry
have these $1/r$ long-range tails. Their presence is the reflection of the Higgs
branch which emerges in the bulk theory in the limit $\mu\to\infty$ \cite{SYnone}.

\section{Parameters of the heterotic \boldmath{$CP(1)$} model from the
bulk theory}
\label{crossterm}
\setcounter{equation}{0}

To derive kinetic terms for two-dimensional fermions in the string
worldsheet theory in the presence of the bulk deformation (\ref{msuperpotbr})
we substitute the fermion zero modes found in  
Sect.~\ref{N1zeromodes} 
in the fermion kinetic terms of the bulk theory (\ref{fermact}). As usual we assume 
the two-dimensional fields to have an adiabatic dependence 
on the worldsheet coordinates.
Note, that the kinetic terms for fields $S^a$, $x_{0i}$, $\chi^a_L$ and $\zeta_L$ 
are not modified. They are still given by Eq.~(\ref{ntwocp}). The reason is that neither
the bosonic string solution (\ref{sna}) nor the fermion zero modes associated with 
unbroken \none supersymmetry are  modified by the deformation (\ref{msuperpotbr}).

Explicitly, the  kinetic terms of the worldsheet theory are 
\beqn
S_{1+1}^{\rm kin}
&=&
 \int d^2 x \left\{2\pi\xi\left[\frac12
\left(\pt_k x_{0i} \right)^2+
\frac12 \, \bar{\zeta}_L \, i
\left(\pt_0+i\pt_3
\right)\, \zeta_L
\right.\right.
\nonumber\\[3mm]
&+& 
\left.
\frac{I_{\zeta}}{2} \, \bar{\zeta}_R \, i
\left(\pt_0-i\pt_3
\right)\, \zeta_R\right]
\nonumber\\[3mm]
&+& 
\beta \left[\frac12 (\pt_k S^a)^2+
\frac{I_{\chi}}{2}\, \chi^a_R \, i(\pt_0-i\pt_3)\, \chi^a_R
+
\frac12 \, \chi^a_L \, i(\pt_0+i\pt_3)\, \chi^a_L
\right.
\nonumber\\[4mm]
&+&
I_{\zeta\chi}\,\chi_R^a\left(\rule{0mm}{6mm}
i(\pt_0-i\pt_3)\,S^a\,(\zeta_R +\bar{\zeta}_R)
+
\left.
\left.
i\varepsilon^{abc} S^b
i(\pt_0-i\pt_3)\,S^c\,(\zeta_R-\bar{\zeta}_R)\right)
\right]
\right\},\nonumber\\[1mm]
\label{kinterm}
\eeqn
where 
$I_{\zeta}$, $I_{\chi}$ and $I_{\zeta\chi}$ are normalization integrals
determined by the profile functions of the fermion zero modes.
The kinetic terms of $x_{0i}$ and $\zeta_L$ are irrelevant;
we present them only for completeness. 
We will consider small and large-$\mu$ limits separately.

\subsection{Small-\boldmath{$\mu$} limit}

To the leading order in $\mu$ 
\beq
I_{\zeta}=I_{\chi}=1\,,
\eeq
i.e the normalization integrals in front of the $\chi^a_R$ and $\zeta_R$ kinetic terms
are still determined by $\beta$.
The only new element is the emergence of the bifermion mixing
 $\chi^a_R\partial_LS^a \zeta_R$ 
which is linear in the deformation parameter $\mu$. 
It can be expressed in terms  of the profile functions as follows: 
\beqn
I_{\zeta\chi}
&=&
\int rdr\left[-\left(\lambda_{t1}\lambda_{+}+\lambda_{t0}\lambda_{-}
\right)\,\frac{\phi_1}{\phi_2}
+\frac{g_2^2}{4}\left(\psi_{t1}\psi_{+}+\psi_{t0}\psi_{-}\right)
\left(1+\frac{\phi_1}{\phi_2}\right)
\right.
\nonumber\\[3mm]
&+&
\left.
\frac{g_2^2}{4}\left(\psi_{s0}\psi_{-}-\psi_{s1}\psi_{+}\right)
\left(1-\frac{\phi_1}{\phi_2}\right)
\right]\,.
\label{Ichizetadef}
\eeqn
Substituting  the leading-order fermion profile functions from Eqs.
(\ref{tzeroorder}) -- (\ref{firstorder}) in Eq. (\ref{Ichizetadef}) we get
\beqn
I_{\zeta\chi}=-\frac{g^2\mu}{2\sqrt{2}}\,\int rdr \left[\frac{g^2(\phi_2^2-\phi_1^2)^2}
{\phi_2^2}
\left(1+\frac12\,f+\frac32 f_3\right)+4g^2(\phi_1^2-\phi_2^2)f_3\right]
+O(\mu^2).\nonumber\\
\label{Ichizetasmu}
\eeqn
where we put $g_1=g_2=g$.

Comparing this with Eq.~(\ref{02o3}) we see that the deformation parameter $\kappa$ is 
\beq
\kappa=I_{\zeta\chi}={\rm const}\,\,g^2\mu +O(\mu^2)\,.
\label{kappasmu}
\eeq
To find the value of the constant in (\ref{kappasmu})
one needs to calculate the overlap integral (\ref{Ichizetasmu}) numerically.
This has not yet been done. What is important is that this constant
is just a number, presumably of order one.

We see that in the small-$\mu$ limit the \ntwoo deformation parameter 
$\kappa$ is proportional to $g^2\, \mu$, in accordance with the
Edalati--Tong suggestion. This parameter (of dimension of mass) determines the mass splitting in the
\ntwo multiplets of the bulk theory upon $\mu$-deformation, see (\ref{omega}).

\subsection{Large-\boldmath{$\mu$} limit}

In this limit the $\lambda$ fields  decouple and the normalization integrals in 
Eq. (\ref{kinterm}) are given by the $\psi$-fermion profile functions,
\beqn
I_{\zeta}
&=&
\frac1{\xi}\int rdr\left(\psi_{s0}^2+\psi_{t0}^2 +\psi_{s1}^2+\psi_{t1}^2\right),
\nonumber\\[3mm]
I_{\chi}
&=&
2g^2_2\int rdr\left(\psi_{+}^2+\psi_{-}^2 \right),
\nonumber\\[3mm]
I_{\zeta\chi}
&=&
\frac{g_2^2}{2}\int rdr\left(\psi_{t1}\psi_{+}+\psi_{t0}\psi_{-}\right).
\label{normint}
\eeqn
All three normalization integrals contain large logarithms
due to the $1/r$ long-range tails of the fermion profile functions. Explicitly,
substituting here Eqs. (\ref{largerpsi0}) and (\ref{largerpsi-})  we get 
\beqn
I_{\zeta}
&=&
2\ln{\frac{m_W}{m_L}} + O(1)\,
\nonumber\\[3mm]
I_{\chi}
&=&
2g_2^2\left[\ln{\frac{m_W}{m_L}} + O(1)\right]\,,
\nonumber\\[3mm]
I_{\zeta\chi}
&=&
\frac{g^2_2\sqrt{\xi}}{2}\left[\ln{\frac{m_W}{m_L}} + O(1)\right]\,,
\label{logs}
\eeqn
where $m_W$ is the gauge field mass (\ref{wmass}) while $m_L$ is the small mass (\ref{lmass})
of the fields $\tilde{q}$ and their fermionic superpartners. The large logarithmic contributions
in (\ref{logs}) come from the integration over $r$ in the domain 
$$1/m_W\ll r\ll 1/m_L\,.$$
Note that the fermion profile functions which have logarithmic behavior in this domain
(see Sect.~8.2) do not produce large logarithms. The corrections to the leading behavior
in (\ref{logs}) come from numerical  constants in the arguments of logarithms which we do not
control in our approximation.

Now we rescale the fields $\zeta_R$ and $\chi_R$ absorbing the logarithms
in $I_\zeta$ and $I_\chi$  in the  
normalization of these fields. Then we arrive at the Lagrangian (\ref{02o3}) with 
\beq
c\kappa=\frac{\kappa}{\sqrt{1+8\frac{\kappa^2}{m_W^2}}}= \frac{m_W}{4}\,,
\label{ckappa}
\eeq
which is equivalent to
\beqn
\kappa
&=&
\frac{m_W}{2\sqrt{2}}\left[1+O\left(1/\ln{\frac{g^2_2\mu}{m_W}}\right)\right], 
\qquad c=\frac{1}{\sqrt{2}}\left[1+O\left(1/\ln{\frac{g^2_2\mu}{m_W}}\right)\right]\,,
\nonumber\\[3mm]
\alpha
&=&
1+O\left(1/\ln{\frac{g^2_2\mu}{m_W}}\right)
\label{kappa}
\eeqn
at large $\mu$. In this regime the worldsheet deformation parameter $\kappa$ is
proportional to the gauge multiplet mass $m_W$. Corrections to the leading behavior here go in inverse powers of the large logarithm.

Summarizing, we found the worldsheet deformation parameter  from the bulk theory
in two limits, small and large $\mu$.
As was stressed in Sect.~5, the deformation of the worldsheet theory is determined by
 the single  dimensionless parameter $\alpha$ (see (\ref{c})), which is the ratio 
of $\kappa$ and the gauge  boson mass.
Our result for this parameter in terms of parameters of the bulk theory reads
\beq
\alpha =2\sqrt{2}\,\frac{\kappa}{m_W}=\left\{
\begin{array}{l}
{\rm const}\,\rule{0mm}{6mm}
 \frac{g^2 \mu}{m_W}\,,\qquad \qquad\;\;\;\;\; \mbox{small}\,\,\,\mu\,,\\[2mm]
1+O\left(1/\ln{\frac{g^2_2\mu}{m_W}}\right)\,, \qquad  \mbox{large}\,\,\,\,\mu\,.
\end{array}
\right.
\label{alpharesult}
\eeq

If we translate this behavior in the behavior of the parameter $\delta$ which enters
the gauged formulation of the \ntwoo $CP(1)$ model (see Sect.~5 and Appendix D for relations between different definitions of the worldsheet deformation parameters) we find that the parameter $\delta$ tends to infinity in the large-$\mu$ limit. Namely, from (\ref{alpharesult})
we get 
\beq
\delta =\frac{\alpha}{\sqrt{1-|\alpha|^2}}=\left\{
\begin{array}{l}
{\rm const}\, \frac{g^2 \mu}{m_W}\,,\qquad \;\;\;\; \mbox{small}\,\,\,\mu\,,\\[3mm]
{\rm const}\,\sqrt{\ln{\frac{g^2_2\mu}{m_W}}}\,, \;\quad  \mbox{large}\,\,\,\,\mu\,.
\end{array}
\right.
\label{deltaresult}
\eeq
The Edalati--Tong suggestion
\cite{Edalati} anticipates
$\delta\sim\mu$
rather than the logarithmic  behavior implied by
(\ref{deltaresult}) at large $\mu$.

As was mentioned in Sect.~2 (see also \cite{SYnone}), ``large $\mu$" here means the values of $\mu$
at the upper limit of  the window (\ref{window}).

As was already explained, the  logarithmic behavior of our results at large $\mu$
is due to   light states with mass $m_L$ . These states are related to the presence of the Higgs branch  in  the bulk theory in the limit $\mu\to\infty$.

\section{Twisted mass in the worldsheet theory}
\label{twistedmass}
\setcounter{equation}{0}

The remainder of this paper is devoted to a more general polynomial
bulk theory deformation presented
by the superpotential (\ref{defsup}). 
Our goal in this section is to prepare for the analysis of polynomial deformations.
Here we will introduce unequal
quark mass terms in the undeformed \ntwo bulk theory and review modifications
that occur in the worldsheet theory due to $m_1\neq m_2$ \cite{SYmon,HT2}.
In Sect.~\ref{deformationsup} we will discuss deformation of the \ntwo bulk theory by the
superpotential (\ref{defsup}).

Thus, let us drop the assumption (\ref{zeromass}) of equal mass terms
for two flavors and introduce a small mass difference $\Delta m$. By shifting the 
adjoint field we can always ensure that 
\beq
m_1+m_2=0\,.
\label{zeroavmass}
\eeq
Without loss of generality we will assume Eq.~(\ref{zeroavmass}) to be satisfied.
With unequal mass terms, the  U(2) gauge group 
of the bulk theory is broken  down to U(1) by the  condensation
of the adjoint scalars, see (\ref{avevm}). The masses of off-diagonal
gauge bosons and off-diagonal fields from the squark matrix $q^{kA}$
(together with their fermion superpartners) get a shift, with splittings proportional to
 $\Delta m$ (we assume $|\Delta m | \ll \sqrt\xi$). 

The Abelian $Z_2$ strings (\ref{znstr}) are  now {\em the only} 
solutions to the first-order string  equations. The family of solutions is discrete.
The global SU$(2)_{C+F}$ group is broken down to U(1) by $\Delta m \neq 0$,
and the moduli space of the non-Abelian string
is lifted. In fact, the vector $ S^a$ gets fixed in two possible positions,
$S^a=(0,0,\pm 1)$. If the mass difference is much smaller than
$\sqrt{\xi}$ the set of parameters  $S^a$ becomes {\em quasi}moduli.

We will outline here derivation of  the  effective two-dimensional theory
on the string worldsheet for unequal  mass terms \cite{SYmon}.
Under the condition $\Delta m \neq 0$ we will still be able to introduce orientational
quasimoduli $S^a$.  In terms of the worldsheet model, unequal mass
terms lead to a shallow potential for the quasimoduli  $S^a$. Let us derive this potential.

To this end we start from the
expression for the non-Abelian string
in the singular gauge (\ref{sna}) parametrized by the moduli $S^a$,
and substitute it in the bulk potential (\ref{pot}). The only
modification we actually have to make is supplementing our
{\em ansatz} (\ref{sna}) by an {\em ansatz} for the adjoint scalar field $a^a$;
the U(1) scalar field $a$ will stay fixed at its VEV,  $a=0$.

At large $r$ the field $a^a$ tends to its VEV
aligned along
the third axis in the color space,
\beq
\langle a^3 \rangle = -\frac{\Delta m}{\sqrt{2}},\;\;\; \Delta m= m_1-m_2,
\label{N2avev}
\eeq
see Eq.~(\ref{avevm}). At the same time, at $r=0$
it must be directed along the vector $S^a$. The reason for this behavior
is easy to understand. The kinetic term for $a^a$  in Eq.~(\ref{model})
contains the commutator term of the adjoint scalar and the gauge potential.
The gauge potential is singular at the origin, as is seen
from  Eq.~(\ref{sna}). This
implies that $a^a$ must be aligned along $S^a$ at $r=0$. Otherwise,
the string tension would become divergent.
The following {\it ansatz} for $a^a$ ensures this behavior:
\beq
a^a=-\frac{\Delta m}{\sqrt{2}}\, \left[\delta^{a3}\, (1-\omega) +S^a\,  S^3\,
\omega\right]\, .
\label{aa}
\eeq
Here we introduced a new profile function $\omega(r)$ which  will be
determined from a minimization procedure \cite{SYmon}.
Note that at $S^a=(0,0,\pm 1)$ the field $a^a$ is given by its VEV,
as expected. The boundary conditions for  the function $\omega(r)$ are
\beq
\omega(\infty)=0\, ,\qquad  \omega(0)=1\, .
\label{bbc}
\eeq
Substituting Eq.~(\ref{aa}) in conjunction with (\ref{sna}) in the bulk potential
(\ref{pot})
we get the potential
\beq
V_{{ CP}(1)}=\beta_{\rm pot}\int d^2 x\, \frac{|\Delta m|^2 }{2} \left(1-S_3^2\right)\, ,
\label{sigpot}
\eeq
where $\beta_{\rm pot}$ is given by  the integral
\beqn
\beta_{\rm pot}
&=&
 \frac{2\pi}{g_2^2} \, \int_0^{\infty}
r\, dr\, \left\{\left(\frac{d}{dr}\omega (r)\right)^2
+\frac{1}{r^2}\, f_{3}^2\,  (1-\omega)^2\right.
\nonumber\\[4mm]
&+&
\left.
g_2^2\left[\frac{1}{2}\,  \omega^2\, \left(\phi_1^2+\phi_2^2\right)
+(1-\omega)\, \left(\phi_1-\phi_2\right)^2\right]\right\}\, .
\label{gamma}
\eeqn
The first and second terms in the integrand come from the kinetic term
of  the adjoint scalar field $a^a$ in (\ref{model}) while the term in the square
brackets comes from the  potential  (\ref{pot}).

Minimizing with respect to $\omega(r)$, with the constraint (\ref{bbc}),
we arrive at
\beq
\omega(r)=1-\frac{\phi_1}{\phi_2} (r)\, .
\label{bobr}
\eeq
This gives
\beq
\beta_{\rm pot} =  \frac{2\pi}{g_2^2}\,= \beta.
\eeq
We see  \cite{SYmon} that the normalization integrals are the same for both, the kinetic
and the potential terms in the worldsheet sigma model, 
$\beta_{\rm pot}=\beta$.
As a result we arrive at  the following effective theory on the string 
worldsheet:
\beq
\label{o3mass}
S_{{ CP}(1)}=\beta \int d^2 x \left\{\frac12 \left(\pt_k
S^a\right)^2 +\frac{|\Delta m|^2}{2}\,\left( 1-S_3^2\right)\right\}\, .
\eeq
This is the only functional form that allows \ntwo completion.\footnote{
Note, that although the global SU(2)$_{C+F}$ is broken by $\Delta m$,
the extended ${\mathcal N}=2$  supersymmetry is not. }
For generic $N$ the potential in the $CP(N-1)$ model was
obtained in \cite{HT2}. 

The $CP(N-1)$ model with the potential (\ref{o3mass}) is nothing but a bosonic 
truncation of the \ntwo two-dimensional sigma model
which was termed the twisted-mass-deformed $CP(N-1)$ model.
This is a generalization of the massless $CP(N-1)$ model which
preserves four supercharges. Twisted chiral superfields in two dimensions
were introduced in \cite{Alvarez} while the twisted mass as an expectation value
of the twisted chiral multiplet was suggested in \cite{HH}. $CP(N-1)$ models
with twisted mass were further studied in \cite{Dorey} and, in particular,
the BPS spectra in these theories were determined exactly.

The fact that we obtain this form shows that our {\em ansatz} is
fully adequate. 
The mass-splitting parameter $\Delta m$ of the bulk
theory exactly coincides with the twisted mass of the worldsheet  model,
\beq
m_{\rm tw}=\Delta m.
\label{twmdeltam22}
\eeq

The $CP(1)$ model (\ref{o3mass}) has two vacua located at 
$S^a=(0,0,\pm 1)$. Clearly these two vacua correspond to two elementary  $Z_2$ strings.

The twisted-mass-deformed $CP(N-1)$ model can be  written as a
strong coupling limit of a U(1) gauge theory \cite{Dorey}, see (\ref{cpg}). With twisted 
masses
of the $n^l$ fields taken into account, the bosonic part of the action
(\ref{cpg}) becomes
\beqn
S_{CP(1)\,{\rm bos}}
& =&
\int d^2 x \left\{
 |\nabla_{k} n^{l}|^2 +\frac1{4e^2}F^2_{kl} + \frac1{e^2}
|\pt_k\sigma|^2+\frac1{2e^2}D^2
\right.
\nonumber\\[3mm]
 &+&    2\left| \sigma-\frac{m_l}{\sqrt{2}}\right|^2 |n^{l}|^2 + iD (|n^{l}|^2-2\beta)
\Big\}\,.
\label{mcpg}
\eeqn
The vacuum expectation values of the $\sigma$ field are determined by the quark mass
terms. For  $N=2$ we have
\beq
\langle \sigma\rangle =\pm\frac{\Delta m}{2\sqrt{2}}\,.
\label{sigmavev}
\eeq

In the limit $e^2\to \infty$ the $\sigma$ field can be eliminated by virtue of an algebraic
equation of motion. For $N=2$ we get
\beq
\sigma=\frac{\Delta m}{4\sqrt{2}\beta}\,(|n^1|^2-|n^2|^2) =\frac{\Delta m}{2\sqrt{2}}\;S_3\,,
\label{sigmap}
\eeq
where we also used (\ref{Sn}). This leads to the potential in (\ref{o3mass}). 

\section{Adding the polynomial deformation superpotential}
\label{deformationsup}
\setcounter{equation}{0}

Now it is time to switch on the 
polynomial deformation superpotential (\ref{defsup}). Classically, this deformation  does not 
spoil the BPS nature of the $Z_2$ strings under consideration. In fact, the string solution remains intact. At $\Delta m\neq 0$ it is still given by Eq.~(\ref{znstr}). 
The reason, as was already mentioned, is that for the deformation superpotential of a special type, with the critical points coinciding  with quark mass terms,
the fields $\tilde{q}$ do not condense in the vacuum. The squark
VEV's are still given by (\ref{qvev}). This ensures the field $\tilde{q}$ to remain 
unexcited on the string solution (at the classical level). The string is made from 
the gauge fields and $q$ fields which have the same mass. Then BPS saturation ensues.

However, the global SU(2)$_{C+F}$ group is broken by  $\Delta m\neq 0$ already
in the undeformed \ntwo theory, see Sect.~\ref{twistedmass}. As a result $S^a$ now become quasimoduli and  
a shallow potential on the moduli space is generated in the effective worldsheet
model, 
see (\ref{o3mass}).
Adding the deformation superpotential (\ref{defsup}) in the bulk theory modifies this potential on the string worldsheet. In this section we will study this modification.

We first focus on modifications of the bosonic potential and 
then restore the heterotic  model in its entirety by calculating the first bosonic term in (\ref{etong}).

From (\ref{pot}) we extract the deformation of the bosonic potential in the bulk theory
\beq
\delta V= g^2\,{\rm Tr}\left|\frac{\pt{\mathcal W}_{3+1}}{\pt\hat{a}}
\right|^2,
\label{deltapot}
\eeq
where the bulk superpotential is determined by (\ref{defsup}) and we take $g_1=g_2=g$ to 
simplify calculations.

As for  the  adjoint field $a^a$ we take the same {\it ansatz} (\ref{aa}) 
as was used in 
Sect.~\ref{twistedmass} for unequal quark mass terms. Substituting it in (\ref{deltapot}) we get
\beq
\delta V_{1+1}=g^2\frac{\pi}{32}\left|\mu\Delta m\right|^2 \int d^2 x
\left(1-S_3^2\right)^2 \int rdr\,\omega^2(2-\omega)^2.
\label{deltaV}
\eeq
The profile function $\omega$ here should be determined via minimization procedure, much
in the same way as it was done in Sect.~\ref{twistedmass} for \ntwo theory.
We consider  separately the cases of small and large $\mu$. The fact that it is the square of
$1-S^2_3$ that enters tells us that 
supersymmetry of the worldsheet model cannot be \ntwot $\!\!$. Then, it must be
\ntwoo$\!\!$.

\subsection{Small-\boldmath$\mu$ limit}
\label{smmuu}

For small $\mu$ we consider (\ref{deltaV}) as a perturbation. To the leading order
in $\mu$ we use the expression (\ref{bobr}) for the profile function $\omega$  obtained 
in the  \ntwo limit. Then the deformation of the worldsheet theory is 
\beq
\delta V_{1+1}=\beta\,\frac{I}{64}\,\frac{g^4|\mu|^2|\Delta m |^2}{m^2_W} \int d^2 x
\left(1-S_3^2\right)^2 ,
\label{deltaVsmu}
\eeq
where $I$ is a dimensionless numerical factor determined by the string profile functions 
\beq
I= m^2_W\int rdr\,\left(1-\frac{\phi_1^2}{\phi_2^2}\right)^2.
\label{I}
\eeq
The bosonic part of the effective theory on the string worldsheet 
is given by the sum of the twisted mass
$CP(1)$ model, Eq.~(\ref{o3mass}), and the deformation potential (\ref{deltaVsmu}).
We see that the points $S^a=(0,0,\pm 1)$ remain to be the vacua of the deformed theory.
The corresponding vacuum energy density vanishes at the classical level. These
vacua describe two $Z_2$ strings of the bulk theory. \ntwoo supersymmetry is not broken
at the classical level.

Now, we can rewrite the potential (\ref{deltaVsmu}) in the form of the   
deformation of the $CP(1)$ model (\ref{mcpg}) in the  gauged formulation, see Eq.~(\ref{etong}). To the leading order 
in $\mu$  the $\sigma$ field
is determined by  Eq.~(\ref{sigma}) obtained in the \ntwot limit. Therefore
we can write (\ref{deltaVsmu}) as 
\beq
\delta V_{1+1}=\int d^2 x\;\frac{8\beta}{m_W^2}\left|\frac{\pt {\mathcal W}_{1+1}}{\pt \sigma}
\right|^2
\label{Trep}
\eeq
with  
\beq
\frac{\pt {\mathcal W}_{1+1}}{\pt \sigma}=\frac{\sqrt{I}}{2\sqrt{2}}\,\frac{g^2\mu}{\Delta m}\,
\left(\sigma^2
-\frac{\Delta m^2}{8}\right).
\label{2dsupsmu}
\eeq

This result is in accordance with the Edalati--Tong suggestion \cite{Edalati}, see 
Sect.~\ref{0,2}. The critical
points of the two-dimensional superpotential are determined by the quark mass
terms, so supersymmetry is not broken in the vacua (\ref{sigmavev}) at the classical level. From the standpoint of the 
bulk theory this means that $Z_2$ strings are BPS saturated. This condition was the motivation behind
the Edalati--Tong suggestion \cite{Edalati}. The coefficient in 
front of the polynomial in $\sigma$ in ${\mathcal W}_{1+1}$
is proportional to $g^2\mu$, much in the same way as for deformation (\ref{msuperpotbr}),
see Eq.~(\ref{kappasmu}). This parameter determines the mass splitting in \ntwo multiplets
of the bulk theory in the small-$\mu$ limit, see  Sect.~\ref{bulk}.

\subsection{Large-\boldmath$\mu$ limit}

In this limit we have to add the deformation (\ref{deltaV}) to the $CP(1)$ model potential 
(\ref{sigpot}) and carry out minimization in order to find the modified
profile function $\omega (r)$.

If $S_3^2$ is very close to unity, $$1-S_3^2\ll m_W/g^2\mu\,,$$ then the deformation in (\ref{deltaV})
still can be considered as a perturbation, much in the same way as it was done   
Sect.~\ref{smmuu}. In this case the profile function $\omega (r)$ stays intact and the result for deformation of the worldsheet theory is still given by Eqs.~(\ref{deltaVsmu}) and (\ref{2dsupsmu}).
However, let us consider the range of $S_3^2$ not too close to unity. In this 
case the deformation (\ref{deltaV})
becomes large at large $\mu$. Minimization with respect to $\omega (r)$ requires
$\omega$ 
to tend to zero in this limit. However, the boundary conditions (\ref{bbc}) tell us that 
$\omega (r) $ cannot vanish for all $r$.

Taking this into account it is natural to assume  the following simple profile for 
$\omega (r)$:
\beq
\omega=
\left\{
\begin{array}{cc}
1, & r<r_0\,, \\[2mm]
0 & r>r_0\,,\\
\end{array}
\right.
\label{omegaprofile}
\eeq
where the parameter $r_0$ should be 
found from minimization. It will turn out to be very close to zero.

Indeed,
\beq
V_{1+1}=\beta \int d^2 x \left\{\frac{|\Delta m|^2}{2}\,\left( 1-S_3^2\right)\,
\ln{\frac{1}{r_0\, m_W}} +
\frac{g^4}{64}\,\left|\mu\Delta m\right|^2 
\left(1-S_3^2\right)^2 \frac{r_0^2}{2}\right\},
\label{2dpot}
\eeq
where the leading logarithmic contribution  comes from the second term in (\ref{gamma}).
The upper limit of the logarithmic integral over $r$ is given by  the inverse gauge boson
mass. At  $r\sim 1/m_W$ the profile function $f_3$ is no longer constant.
It cuts the logarithmic integration.
Minimizing with respect to $r_0$ we obtain
\beq
r_0^2\sim \frac{1}{g^4|\mu|^2\,\left( 1-S_3^2\right)}\,.
\label{r0}
\eeq
We see that $r_0$ is very close to zero, indeed. It is determined by the mass
of the adjoint fields $g^2\mu$ which tends to infinity at $\mu\to\infty$.

Substituting this back in Eq.~(\ref{2dpot}) we get 
\beq
V_{1+1}=\beta \int d^2 x \frac{|\Delta m|^2}{2}\,\left( 1-S_3^2\right)\,
\left[\ln{\frac{g^2|\mu|}{ m_W}}+\frac12\ln{\left( 1-S_3^2\right)}\right].
\label{V}
\eeq
This is our final result for the bosonic potential in the \ntwoo worldsheet theory on the 
string. Undetermined non-logarithmic corrections to this leading logarithmic expression
come from 
the second term in (\ref{2dpot}), other terms in (\ref{gamma}), as well as from improvements of the simple step-function profile (\ref{omegaprofile}).

If $S^2_3$ is not too close to unity, we can neglect the second term in (\ref{V}). We see that the main effect of the bulk deformation is the modification of the twisted mass parameter of the $CP(1)$ model. Now, expressed in terms of the bulk parameters 
it acquires a dependence on $\mu$ and $\xi$. Instead of the simple expression
(\ref{twmdeltam22}) we now get an ``amplified" twisted mass,
\beq
m_{\rm tw}=\Delta m\,\sqrt{\ln{\frac{g^2|\mu|}{ m_W}}}\,.
\label{mtw}
\eeq
The twisted mass becomes logarithmically large as we increase $\mu$. The worldsheet theory coupling becomes weaker. 

If $S^2_3$ is close to unity, $ m_W/g^2\mu\ll1-S_3^2\ll 1$  we  keep the second term
in (\ref{V}) as a correction to the leading twisted mass term. This term also has zeros 
at $S^a=(0,0,\pm 1)$. Therefore,  the vacua  of the worldsheet theory
remain intact and supersymmetry is not broken at the classical level. However, the potential becomes nonpolynomial in $S_3$.

We can rewrite our results in terms of the gauged formulation. We have
\beqn
S_{1+1}
& =&
\int d^2 x \left\{
 |\nabla_{k} n^{l}|^2 +\frac1{4e^2}F^2_{kl} + \frac1{e^2}
|\pt_k\sigma|^2+\frac1{2e^2}D^2
\right.
\nonumber\\[3mm]
 &+&    2\left |\sigma-\frac{m_{\rm tw}}{2\sqrt{2}}\right|^2 |n^{1}|^2 +
2\left|\sigma+\frac{m_{\rm tw}}{2\sqrt{2}} \right |^2 |n^{2}|^2
+ iD (|n^{l}|^2-2\beta)
\nonumber\\[3mm]
 &+&
\left.
\beta\frac{|\Delta m|^2}{4}\left|\left(1-\frac{8\sigma^2}{m_{\rm tw}^2}\right)\,
\ln{\left(1-\frac{8\sigma^2}{m_{\rm tw}^2}\right)}\right|
\right\}\,,
\label{1+1g}
\eeqn
where the twisted mass is given by (\ref{mtw}). The last term in the potential here is small and 
can be considered as a perturbation. Then the equation of motion for the $\sigma$ 
field gives,
to the leading order,
\beq
\sigma=\frac{m_{\rm tw}}{4\sqrt{2}\beta}\,(|n^1|^2-|n^2|^2) =\frac{ m_{\rm tw}}{2\sqrt{2}}\;S_3\,.
\label{sigmamod}
\eeq
Being substituted in the last term in (\ref{1+1g}) it reproduces the last 
logarithmic potential term in (\ref{V}).

The last  term in (\ref{1+1g}) can be written in the form (\ref{Trep}) with the 
\ntwoo superpotential
\beq
\frac{\pt {\mathcal W}_{1+1}}{\pt \sigma}=
m_W\frac{\Delta m}{4\sqrt{2}}\,\left[\left(1-\frac{8\sigma^2}{m_{\rm tw}^2}\right)\,
\ln{\left(1-\frac{8\sigma^2}{m_{\rm tw}^2}\right)}\right]^{1/2}\,.
\label{2dsuplmu}
\eeq
This superpotential is non-polynomial and does not satisfy conjecture 
(\ref{etconjecture}). It depends on $\mu$ logarithmically via the twisted mass (\ref{mtw}).
Its critical points are determined by the twisted mass and coincide with the vacua of
the theory (\ref{1+1g})
\beq
\langle\sigma\rangle =\pm\frac{m_{\rm tw}}{2\sqrt{2}}\,.
\label{sigmamodp}
\eeq
The vacua of the deformed theory are modified as compared with 
the \ntwot case, see Eq.~ (\ref{sigmavev}).

We would like to stress the following: the most important impact of the \ntwo breaking polynomial deformation of the bulk theory at large $\mu$
is the logarithmic dependence of the worldsheet twisted mass (\ref{mtw}) 
on the ratio $\mu/\sqrt\xi$. In the \ntwo limit the dependence of $m_{\rm tw}$ on the 
nonholomorphic
FI parameter $\xi$ is forbidden \cite{SYmon,HT2}. This is no longer true as we break \ntwo
supersymmetry  down to \none in the bulk. As a result the twisted mass term 
becomes large forcing the string orientational vector $S^a$ to point towards 
the north or south 
poles of $S_2={\rm SU}(2)/{\rm U}(1)$. This means that the string becomes more ``Abelian'' as we increase $\mu$.
This is in accord with ``Abelianization'' of the bulk theory. As was mentioned in 
Sect.~\ref{bulk}, the 
$\mu$ deformation splits adjoint scalar multiplet giving large masses to $a$ and $a^3$
components while the masses of the $a^{1,2}$ components 
are still determined  by  $m_W$ 
(and $\Delta m$).

Once $S_3^2$ is close to unity this effect is partly washed out in the worldsheet theory
by the logarithmic correction in (\ref{V}).

Note that small variations of the polynomial deformation
(shifting the critical points from $m_{1,2}$) ruin the BPS saturation of the classical string solutions. The worldsheet model in this case must break supersymmetry already at the classical level. This means that there is no protection against spontaneous supersymmetry breaking even in the
case when the critical points coincide with $m_{1,2}$. Classically SUSY is unbroken,
the breaking can (and does) occur at the quantum level.

\section{Conclusions}
\label{conclu}

In this paper we continue studies of non-Abelian strings
in the U$(N)$ bulk theories with $N$ flavors. If the bulk theory is \ntwo supersymmetric,
the string is BPS saturated and
the low-energy theory on the string worldsheet has \ntwot supersymmetry.
The worldsheet model splits into two completely disconnected
sectors: (i) noninteracting theory of two translational and
four supertranslational moduli; and (ii) supersymmetric $CP(N-1)$ model
describing interactions of orientational and superorientational moduli.

We start from deforming the \ntwo bulk theory by introducing
deformations (of a special type), which preserve \none in the bulk. The string solution at the classical level remains BPS saturated. Normally, this would imply conservation of two supercharges on the string worldsheet. Previously it was believed, however, that
worldsheet supersymmetry gets an ``accidental" enhancement.
This is due to the facts that ${\mathcal N} =(1,1)$ SUSY
is automatically elevated up to \ntwot on $CP(N-1)$ and, at the same time,
there are no ``heterotic" \ntwoo generalizations of the bosonic $CP(N-1)$ model. 

Edalati and Tong noted that the target space is in fact
$CP(N-1)\times C$ rather than $CP(N-1)$.
If two fermionic moduli from the first sector (see above)
become coupled to moduli from the second sector,
one can built a heterotic \ntwoo model with the $CP(N-1)$
target space for the bosonic moduli. They suggested a general structure of such model
(in the gauged formulation), and a particular formula
relating deformation parameters in the bulk with those on the worldsheet.
Later Tong argued that \ntwoo supersymmetry of the heterotic model
is spontaneously broken at the quantum level. 

Our task was a direct derivation of the string worldsheet model from the bulk theory 
with  \none and the superpotential (\ref{msuperpotbr}) or (\ref{defsup}), including
the relation between the deformation parameters in the bulk
and on the worldsheet. The model we obtain follows the general pattern of Edalati and Tong.
Both, the O(3) formulation and the geometric formulation  which we use
in this paper 
instead of the gauged formulation
serve well our original goal and allowed us to find the full solution.
The Edalati--Tong suggestion as to how
the bulk and worldsheet deformation parameters must be related to each other
turns out to be true only for small values of $\mu$.
For large deformations our expressions are different. 

As was mentioned more than once,
deformation of the worldsheet theory is determined by a single {\em dimensionless}
parameter $\alpha$ (or $\gamma$, see (\ref{gamal})). Our result for this 
parameter is given in Eq.~(\ref{alpharesult}).  The
only dimensional (mass) parameter of  the worldsheet theory  $\Lambda_{CP(1)}$ is generated
by nonperturbative effects in two dimensions. In particular, the \ntwoo deformation does not
involve any new mass parameters.

We derived the heterotic \ntwoo model with the $CP(N-1)$ target space for 
bosonic fields in the geometric formulation (see Eq.~(\ref{cpn-1g})).
This representation
turns out to be very convenient for proving
spontaneous breaking of SUSY at small $\mu$. The vacuum energy density
is shown to be proportional to the square of the bifermion condensate.
Spontaneous breaking for arbitrary values of
deformation parameter (and large $N$) will be proven in \cite{ACC}.

\section*{Acknowledgments}
We are grateful to Stefano Bolognesi, Adam Ritz, David Tong,
and Arkady Vainshtein
for stimulating discussions.
We would like to thank  P. Bolokhov for pointing out to us
the omission of the gauge field contribution in Eqs.~(8.3) and (8.4)
of the original version of the manuscript.
This work  is supported in part by DOE grant DE-FG02-94ER408. 
The work of A.Y. was  supported 
by  FTPI, University of Minnesota, 
by RFBR Grant No. 06-02-16364a 
and by Russian State Grant for 
Scientific Schools RSGSS-11242003.2.

\section*{Appendices}

\addcontentsline{toc}{section}{Appendices}

\renewcommand{\thesubsection}{A}
\subsection*{A. Euclidean  notation}

\renewcommand{\theequation}{A.\arabic{equation}}
\setcounter{equation}{0}

As was mentioned, in Sects. 2-5 and 7-11 we use 
 a formally Euclidean notations, e.g.
\beq
F_{\mu\nu}^2 = 2F_{0i}^2 + F_{ij}^2\,,
\eeq
and 
\beq
(\partial_\mu a)^2 = (\partial_0 a)^2 +(\partial_i a)^2\,,
\eeq
 etc.
This is appropriate, since we mostly consider
static (time-independent)
field configurations, and $A_0 =0$. Then the Euclidean action is
nothing but the energy functional. 

Then, in the fermion sector  we have to define 
 the Euclidean matrices
\beq
(\sigma_{\mu})^{\alpha\dot{\alpha}}=(1,-i\vec{\tau})_{\alpha\dot{\alpha}}\,,
\eeq
and 
\beq
(\bar{\sigma}_{\mu})_{\dot{\alpha}\alpha}=(1,i\vec{\tau})_{\dot{\alpha}\alpha}
\,.
\eeq
 Lowing and raising
 of the spinor indices
is performed by
virtue of the antisymmetric tensor defined as
\beqn
\varepsilon_{12}=\varepsilon_{\dot{1}\dot{2}}=1\,,
\nonumber\\[2mm]
\varepsilon^{12}=\varepsilon^{\dot{1}\dot{2}}=-1\,.
 \eeqn
 The same raising and lowering convention applies to the flavor SU(2)$_{R}$
 indices $f$, $g$, etc.

When the contraction of the spinor indices is assumed inside the parentheses 
we use the 
following notation:
\beq
(\lambda\psi)\equiv \lambda_{\alpha}\psi^{\alpha},\qquad 
(\bar{\lambda}\bar{\psi})\equiv \bar{\lambda}^{\dot{\alpha}}\bar{\psi}_{\dot{\alpha}}
\,.
\eeq

The bar (overline) denotes Hermitian conjugation both in four and two dimensions.

\renewcommand{\theequation}{B.\arabic{equation}}
\setcounter{equation}{0}

\renewcommand{\thesubsection}{B}

\subsection*{B. Two-dimensional Minkowski notation}

\renewcommand{\theequation}{B.\arabic{equation}}
\setcounter{equation}{0}

\beq
\gamma^0=\gamma^t =\sigma_2\,,\quad \gamma^1=\gamma^z = i\sigma_1
\,,\quad
\gamma^0\gamma^1= \sigma_3\,.
\eeq
\beq
g^{\mu\nu} = {\rm diag }\{+1,\, -1\}\,.
\eeq
\beq
\psi =\left(\begin{array}{cc}
\psi_R \\
\psi_L
\end{array}
\right),\quad
\bar \psi =\psi^\dagger\gamma^0\,,\quad \bar \theta =\theta^\dagger\gamma^0\,.
\eeq
\beq
\partial_L =\frac{\partial}{\partial t} +\frac{\partial}{\partial z}\,,\qquad
\partial_R =\frac{\partial}{\partial t} - \frac{\partial}{\partial z}\,.
\eeq

\subsection*{C. Witten index}

\renewcommand{\theequation}{C.\arabic{equation}}
\setcounter{equation}{0}

Witten was the first to calculate ${\rm Tr}\left(-1\right)^F$ in $CP(N-1)$
models, see Sect. 10 of \cite{WI}. It turns out to coincide with the Euler
characteristic of the target space, i.e. $I_W= N$  for  $CP(N-1)$. For $CP(1)$
treated in a small box, as in \cite{WI},
we have two vacua (e.g. the north and south poles of the sphere), both of them bosonic.
Hence $I_W= 2$. Introduction of an extra field $\zeta_R$, as in Eq.~(\ref{AAone}),
splits each of these vacua in two, one bosonic and one fermionic
(since the $\zeta_R$ zero level can be either filled or empty).
Thus, in the heterotic \ntwoo model based on $CP(1)$
\beq
I_W = 0.
\eeq
This is in full agreement with the consideration carried out in Sect.~\ref{geom}
where it was proven that supersymmetry is spontaneously broken at small but
nonvanishing values of $\gamma$. A proof for generic values 
of the deformation parameter but large $N$ is presented
in \cite{ACC}.

\subsection*{D. Worldsheet deformation parameters}

\renewcommand{\theequation}{D.\arabic{equation}}
\setcounter{equation}{0}

For convenience in this Appendix we summarize different definitions of the 
worldsheet deformation parameter.

In O(3) sigma model formulation given in Sect.~5 \ntwoo deformation parameter $\alpha$ is 
related to
parameters which enter the action  (\ref{02o3}) as 
\beqn
\rule{0mm}{9mm}
\alpha &\equiv& \frac{2\sqrt{2}\kappa}{m_W} = \frac{2\sqrt{2}\kappa}{g_2\sqrt{\xi}}\, ,
\nonumber\\[3mm]
c^2&=&\frac{1}{1+|\alpha|^2}\,.
\label{alphakappac}
\eeqn

Its relation to the deformation parameter $\gamma$ which is used for the geometric formulation
of \ntwoo $CP(1)$ model in Sect.~6 is the following:
\beq
\gamma=\sqrt{\beta}\,\frac{\alpha}{\sqrt{1+|\alpha|^2}}.
\eeq

In the gauged formulation of the model (see Sect.~4) it is convenient to use another 
parameter $\delta$,
\beq
\delta=\frac{\alpha}{\sqrt{1-|\alpha|^2}}.
\eeq

In the limit of small and large $\mu$ we have
\beq
\alpha =2\sqrt{2}\,\frac{\kappa}{m_W}=\left\{
\begin{array}{l}
{\rm const}\, \frac{g^2 \mu}{m_W}\,,\qquad \qquad\;\;\; \mbox{small}\,\,\,\mu\,,\\[2mm]
1+O(1/\ln{\frac{g^2_2\mu}{m_W}})\,, \qquad  \mbox{large}\,\,\,\,\mu\,,
\end{array}
\right. 
\label{alpharesultp}
\eeq
and
\beq
\delta =\frac{\alpha}{\sqrt{1-|\alpha|^2}}=\left\{
\begin{array}{l}
{\rm const}\, \frac{g^2 \mu}{m_W}\,,\qquad \;\;\;\; \mbox{small}\,\,\,\mu\,,\\[2mm]
{\rm const}\,\sqrt{\ln{\frac{g^2_2\mu}{m_W}}}\,, \;\quad  \mbox{large}\,\,\,\,\mu\,.
\end{array}
\right.
\label{deltaresultp}
\eeq

\end{document}